\documentclass[a4paper,fleqn,usenatbib,useAMS]{mnras}

\usepackage{graphicx}	
\usepackage{amsmath}	
\usepackage{amssymb}	
\usepackage{multicol}        
\usepackage{bm}		
\usepackage{pdflscape}	

\usepackage[T1]{fontenc}
\usepackage{ae,aecompl}

\title[r-process retainment]{Retainment of r-process material in dwarf galaxies}
\author[P. Beniamini, I. Dvorkin $\&$ J. Silk]{Paz Beniamini$^{1,2}$, Irina Dvorkin$^{3,4}$, Joe Silk$^{4,5}$
	\\
	$^1$Department of Physics, The George Washington University, Washington, DC 20052, USA \\
	$^2$Astronomy, Physics and Statistics Institute of Sciences (APSIS)\\
	$^3$Max Planck Institute for Gravitational Physics (Albert Einstein Institute), Am M\"{u}hlenberg 1, Potsdam-Golm, 14476, Germany\\
	$^4$Institut d'Astrophysique de Paris UMR 7095 Universit\'{e}
	Pierre et Marie Curie-Paris 06; CNRS 98 bis bd Arago, 75014 Paris, France \\
	$^5$Department of Physics and Astronomy, The Johns Hopkins University, Baltimore MD21218 USA}

\begin{document}
	\label{firstpage}
	\pagerange{\pageref{firstpage}--\pageref{lastpage}}
	\maketitle
	
	\begin{abstract}
		The synthesis of $r$-process elements is known to involve extremely energetic explosions. At the same time, recent observations find significant $r$-process enrichment even in extremely small ultra-faint dwarf (UFD) galaxies. This raises the question of retainment of those elements within their hosts. We estimate the retainment fraction and find that it is large $\sim 0.9$, unless the $r$-process event is very energetic ($\gtrsim 10^{52}$erg) and / or the host has lost a large fraction of its gas prior to the event. We estimate the $r$-process mass per event and rate as implied by abundances in UFDs, taking into account imperfect retainment and different models of UFD evolution. The results are consistent with previous estimates \citep{Beniamini2016} and with the constraints from the recently detected macronova accompanying a neutron star merger (GW170817). We also estimate the distribution of abundances predicted by these models. We find that $\sim 0.07$ of UFDs should have $r$-process enrichment. The results are consistent with both the mean values and the fluctuations of [Eu/Fe] in galactic metal poor stars, supporting the possibility that UFDs are the main 'building blocks' of the galactic halo population.
	\end{abstract}
	
	\section{Introduction}
	The  origin of $r$-process elements is a long-standing mystery in chemical evolution modelling \citep{burbidge1957RvMP,cameron1957,Thielemann2017}. The recent gravitational wave event, GW170817 \citep{GW170817} and the identification of a binary neutron star (NS) merger as an $r$-process site \citep{multiGW170817,2017arXiv171005443D,2017arXiv171005858P} lends strong credence to the idea that binary NS mergers are the dominant source of $r$-process elements in the universe as fist suggested by \cite{Lattimer1974,eichler1989Nature} and enables us to compute with more confidence the rate and yield of $r$-process events.
	
	Another fairly recent development has been the measurement of $r$-process elemental abundances (for example [Eu/H]) in extreme metal-poor environments, such as galactic halo stars \citep{Roederer2017}, and, notably, ultra faint dwarf (UFD) galaxies \citep{ji2016Nature,roederer2016}. One UFD, Reticulum II, has been found to have a large excess of $r$-process elements, while all other UFDs have strong upper limits on their $r$-process abundances \citep{ji2016,roederer2014MNRAS}. As was shown in \cite{Beniamini2016}, this is strongly suggestive of enrichment by a single event and enables to break the degeneracy inherent to observations of Milky Way stars, between the rate of $r$-process events and the amount of mass produced per event. Both the implied rate and mass per event are consistent with GW170817, as well as with constraints from observations of short gamma ray burst (GRB) rates and mass ejections from NS merger simulations.

	The low escape velocity of UFDs \citep{Walker2015}, as well as the fact that UFDs are composed of old stellar material (formed within the first Gyr of the galaxys' evolution, \citealt{Brown2014,Weisz2015}) have initially cast some doubt on the association of a NS merger as the origin of the $r$-process material in Reticulum II \citep{Bramante2016}. However, as was shown by \cite{Beniamini2016a}, realistic distributions of orbital parameters and initial kicks received by double neutron star systems \citep{BP2016,Tauris2017} imply that a significant fraction (up to $\sim 60\%$) of double neutron star systems should be able to remain confined even in UFDs as well as merge within less than a Gyr after formation. Even if binary neutron stars do not escape their host galaxies, some of the $r$-process material ejected during their mergers with vast amounts of energy ($\gtrsim 10^{50}$erg) could still escape. If this fraction is significant, this could strongly affect the implied mass of $r$-process material that has to be produced in a single event in order to match the observed abundances.
	
	These results have motivated us to take a more detailed look at how inhomogeneous $r$-process enrichment may occur in ultra faint dwarfs as a consequence of ejecta from rare events (such as, but not restricted to, binary NS mergers), that take place in dwarf galaxies with low escape velocities. There is a general consensus, both theoretical  \citep{Griffen2016} and observational \citep{Frebel2010Natur}, that UFDs are remnants of a large population of dwarfs that long ago dissolved to form the galactic halo stellar population. Therefore, understanding the typical enrichment of $r$-process and iron in those galaxies, could be crucial also to our understanding of metal poor halo stars.
	
	Here we explore simple models for enrichment of $r$-process elements from a very small number of $r$-process events in gas-rich dwarf galaxy precursors of today's ultra faint galaxies. We focus mainly on binary NS mergers, but the results are easily applicable to any other potential $r$-process sources that have been suggested in the literature, such as long GRBs \citep{metzger2008ApJ,vlasov2014MNRAS} or peculiar supernovae (SNe) \citep{suzuki2005ApJ,fryer2006ApJ,Winteler12,nishimura2015ApJ,Moesta15}. Furthermore, these calculations also apply to retainment of iron from core collapse SNe. In what follows, we model the retainment of $r$-process ejecta and calculate the retainment fraction. We include different mass loss models (such as SN-driven mass loss and active galactic nucleus, AGN, driven outflows) as well as iron enrichment due to SNe. For a Reticulum II - like galaxy, we show that unless a significant ($\sim 90\%$) amount of gas has been removed from the galaxy prior to the $r$-process event and/or unless the $r$-process explosion was extremely energetic ($\sim 10^{52}$erg), the retainment fraction of $r$-process materials is large (with typical values around $\sim 0.9$). Furthermore we evaluate the probability of finding single $r$-process enrichment events in dwarf galaxies, and the typical expected iron and $r$-process abundances in such galaxies. These are consistent with observations of galactic metal-poor stars.

	The paper is organized as follows. We begin in \S \ref{sec:rprocejeta} with a brief description of the properties of $r$-process ejecta from compact mergers as constrained by theory and observations. We then discuss the spreading of $r$-process matter in its environment in \S \ref{blastwavephases} with specific implications to Ret II and GW170817. In \S \ref{sec:retainmentfrac} we describe our method for calculating the retainment of an ejecta of an explosion with given energy within its host galaxy. In \S \ref{sec:gasloss} we explore different models for the gas loss in UFDs and its implications on the observed abundances of $r$-process elements. We then use these models to estimate the amount of $r$-process mass created by one event as well as the rate of these events in \S \ref{sec:likelihood}. We address the overall expected distributions of enrichment patterns in UFDs in \S \ref{sec:rareRetII} and explore the dependence of our analysis on the model parameters in \S \ref{sec:varying}. Finally, we discuss some implications of this work and conclude our results in \S \ref{sec:discuss}.
	
	\section{$r$-process ejecta from compact mergers}
	\label{sec:rprocejeta}
	$r$-process ejecta is thought to take place in two different sites: NS-NS and NS-black hole (BH) mergers (see \cite{Metzger2017} for a recent review).
	\begin{itemize}
		\item {\bf Dynamical ejecta} This is material that becomes unbound from the merging binary due to tidal forces during the merger  \citep{Rosswog1999}. It is ejected on a dynamical time-scale (which is of the order of milliseconds). This component exists for NS-BH only if the BH has relatively low mass and / or is spinning rapidly (otherwise the NS could be completely swallowed by the BH). For a NS-NS merger, this component depends critically on the immediate product of the merger (BH or hyper-massive NS). The masses of the dynamical ejecta typically span the range $10^{-3}-10^{-2}M_{\odot}$ with velocities of order the escape velocity, $0.2-0.3c$ (corresponding to kinetic energies of $E_{kin}\sim 4\times 10^{49}-2\times 10^{51}$erg). For BH-NS mergers the masses increase to $0.1M_{\odot}$ (and the velocities are similar). The high velocity component of the dynamically ejected material is squeezed into polar directions and produces the so-called 'blue macronova' \citep{Oechslin2007,hotokezaka2013PRD,sekiguchi2015PRD}. This material is mainly composed of elements with $A<140$, so that the opacity remains sufficiently low and the luminosity  peaks early enough \citep{Kasen2017}. Some of the tidally ejected material will be ejected closer to the equatorial plane. This material may be more lanthanide-rich and contribute partially to the 'red macronova'.
		\item {\bf Disk wind ejecta} During the merger an accretion disc is formed. At early times after its formation, the disk is hot and emits many neutrinos. These neutrinos deposit their energy at larger radii and drive an outflow of material from the disk \citep{Perego2014,Just2015}. This component is expected to be small if the merger immediately produces a BH, but can become very significant if the central NS survives for longer than $\sim 50$msec. The mass ejected in this wind is of order $10^{-2}-10^{-1}M_{\odot}$. It is ejected at velocities of $0.05-0.1c$  (thus corresponding to kinetic energies of $E_{kin}\sim5\times 10^{49}-3\times 10^{51}$erg) and is trailing behind the dynamical ejecta. This material is rather isotropic and is expected to be the main contributor to the 'red macronova'.
	\end{itemize}
	The overall kinetic energy of the merger ejecta is therefore expected to be between $E_{\rm kin}=10^{50}-5\times 10^{51}$erg. We note also that these expected theoretical components described here, match well with the recent observations of GW170817 \citep{Covino2017}. In particular the temperature and luminosity evolution of the macronova associated with GW170817 suggest an expansion velocity of $0.1c-0.3c$ \citep{Kasliwal2017}. Combining this with the estimated ejecta mass of $\sim 0.05M_{\odot}$ we find a kinetic energy of $E_{\rm kin}\sim 5\times 10^{50}-4\times 10^{51}$erg for the $r$-process ejecta in GW170817. Future observations will enable us to narrow down the full distribution of energetics and $r$-process abundances associated with the different ejecta components \citep{Rosswog2017,Cote2017}.

	\section{Spread of $r$-process ejecta}
	\label{blastwavephases}
	The $r$-process ejecta expands into its environment, undergoing a similar evolution to that of a supernova remnant (see also \cite{Montes2016}). Initially the ejecta is expanding freely (at its initial velocity). This stage lasts until the ejecta has collected an interstellar matter (ISM) mass which is of the order of its initial mass. For a spherical ejecta this occurs at $r=1.5 \times 10^{18} M_{ej,-2}^{1/3} n_0^{-1/3}$cm, where $M_{ej}$ is the ejecta mass (in units of $10^{-2}M_{\odot}$) and $n_0$ is the particle density (in $\mbox{cm}^{-3}$). At this stage, the material starts decelerating, following the self-similar Sedov-Taylor expansion. Here $r(t)\propto (\frac{E}{m_pn_0})^{1/5}t^{2/5}$ (where $E$ is the kinetic energy of the ejecta and $m_p$ the proton mass). This stage lasts up until $t_{TR}$, when the decrease in temperature due to radiative cooling overcomes the temperature decrease due to adiabatic expansion \citep{Ostriker1988}. After an intermediate stage between $t_{TR}$ and $\sim 2t_{TR}$ \citep{Haid2016}, the blast wave enters the pressure driven phase \citep{Cox1972}, in which the radius evolves as $r\propto t^{2/7}$. It is typically during this stage that the blast wave reaches its maximum radius, obtained when the shock velocity becomes comparable to the velocity dispersion of the environment, $\sigma_v$. For a homogeneous medium, the transition time depends somewhat on the ionization of the medium and is approximately $t_{TR}\approx 4n_0^{-0.53}\times 10^4$yrs \citep{Haid2016}. We can now estimate the maximum radius (denoted as $r_{\rm fade}$) as
	\begin{equation}
	\label{eq:rfade}
	r_{\rm fade}=70 \bigg(\frac{\sigma_v}{4 \mbox{km s}^{-1}}\bigg)^{-0.4} E_{51}^{0.28} n_0^{-0.365}\mbox{parsec}
	\end{equation}
	where $E_{51}\equiv E/10^{51}\mbox{erg}$.
	Specifically, for a velocity of $\sigma_v=4 \mbox{km s}^{-1}$ (comparable to the velocity dispersion in Ret II and other ultra-faint dwarves), an energy of $10^{51}$erg and for $n_0=1\mbox{cm}^{-3}$, this leads to a final radius of $\approx 70$pc which is slightly more than twice the half-light radius of Ret II. This suggests that some of the $r$-process material created in any given explosion may be ejected outside of its host galaxy.

	Recently, \cite{Macias2016} have used a similar approach to find the minimum amount of mass into which $r$-process material can be injected. This in turn provided them with a maximum abundance that can be associated with a given $r$-process mass that is created in the explosion, and led them to conclude that $r$-process events need to synthesize a minimum of $10^{-3.5}M_{\odot}$ of $r$-process material per event, in order to be able to explain the observed abundances of extremely metal-poor Galactic halo stars.

	Finally, it is interesting to apply these estimates to the recent detection of $r$-process elements in the macronova accompanying GW170817. Observations of the accompanying gamma-ray burst (GRB) afterglow suggest that the environment surrounding the merger has a density $n\approx 10^{-4}-10^{-2}\mbox{cm}^{-3}$ \citep{Margutti2017}. These estimates are also consistent with limits from the mass of neutral hydrogen \citep{Hallinan2017} that find $n<0.04\mbox{cm}^{-3}$. The velocity dispersion in the galaxy is estimated at $\sim 200\mbox{km s}^{-1}$ \citep{Im2017}. Plugging these numbers into Eq. \ref{eq:rfade} we find that $r_{\rm fade}\approx 80-400$pc. Given that the half light radius of the host galaxy, NGC4993 is $R_{1/2}=13$kpc \citep{Abbott2017,Hjorth2017}, the vast majority of the $r$-process material created in GW170817 should be retained by the galaxy. The amount of mass into which this material is injected is approximately
	\begin{eqnarray}
	& M_{\rm fade}=(4\pi/3) m_p n_0 r_{\rm fade}^3 \approx \nonumber \\ & 640  \bigg(\frac{\sigma_v}{200 \mbox{km s}^{-1}}\bigg)^{-1.2}n_{-3}^{-0.1} E_{51}^{0.84} M_{\odot}
	\end{eqnarray}
	where $n_{-3}\equiv n/10^{-3} \mbox{cm}^{-3}$. In particular for $n\approx 10^{-4}-10^{-2}\mbox{cm}^{-3}$ we find $M_{\rm fade}\approx 500-800 M_{\odot}$ (notice that the dependence on the poorly constrained density is very weak). The amount of Eu produced by this event is estimated at $1.2\times 10^{-4} \lesssim m_{\rm Eu} \lesssim 5.3\times 10^{-4}M_{\odot}$ for $A_{\rm min}=90$ and $2.5\times 10^{-5} \lesssim m_{\rm Eu} \lesssim 1.1\times 10^{-4}M_{\odot}$ for $A_{\rm min}=69$ (where $A_{\rm min}$ is the minimum atomic number of the synthesized $r$-process elements, see discussion at the end of \S \ref{sec:likelihood} for more details). These masses correspond to an abundance of $2.6\lesssim \mbox{[Eu/H]} \lesssim 3.4$ for $A_{\rm min}=90$ (or $2.2\lesssim \mbox{[Eu/H]} \lesssim 2.5$ for $A_{\rm min}=69$). 
	These values may be decreased significantly if turbulent mixing occurs, whereas additional $r$-process events in the region could increase this abundance.
	
	\section{Calculation of retainment fraction}
	\label{sec:retainmentfrac}
	We outline here our method for calculating the average fraction of mass that is retained when an $r$-process event of energy $E$ takes place in a given UFD galaxy, characterized by its half light radius $r_{1/2}$ and velocity dispersion $\sigma_v$. The calculation does not assume a specific model for the $r$-process mechanism, and relies only on the total energy of the explosion. Since core-collapse SNe (ccSNe) have similar energies to neutron star mergers' ejecta, similar retainment values are expected for either mechanism. Furthermore, these retainment values would be applicable to any other material that is synthesized in SNe, such as iron.

	The method presented here is generic, but for the sake of concreteness we consider the same sample of UFD galaxies studied by \cite{Beniamini2016}. It includes five UFDs: Reticulum II, Segue I, Segue II, Coma Berenices and Ursa Major II. All of these are composed of old stellar material \citep{Simon2007} and have significantly fewer stars then naively expected from their halo masses, indicating significant mass loss during their evolution.
	\cite{Walker2009} show that a good approximation for the distribution of the halo mass in dwarf galaxies is
	\begin{equation}
	M_h(r)=\frac{5r_{1/2} \sigma_v^2(r/r_{1/2})^3}{G\bigg[1+(r/r_{1/2})^2 \bigg]}\:.
	\end{equation} 
	As a fiducial model we assume here that the initial gas density follows the same distribution as the dark matter and that the total initial gas mass is $M_{g,0}=M_h(r<r_{90})/6$ where $r_{90}\approx 3.71 r_{1/2}$ (models with lower values for the initial gas mass are explored in \S \ref{sec:varying}). Since, as mentioned above, UFDs have lost most of their gas throughout their evolution, it is plausible to assume that the gas mass at the time of the $r$-process explosion, $M_g$, is significantly smaller than the initial gas mass $M_g \ll M_{g,0}$. At the same time, there must be sufficient gas mass that remains confined to the galaxies in order to create the total stellar mass, $M_*$, that resides in them. We conservatively assume here that a gas mass of at least $M_f\approx 10M_*$ is required in order to eventually be reprocessed into the observed stars. Alternatively, if the duration of star formation in UFDs is $\ll 1$Gyr, such that all the gas is re-processed into stars at an earlier stage, $M_f$ could be much smaller, down to the gas mass in UFDs at the present age. The latter may be as low as $10M_{\odot}$ \citep{Spekkens2014}. We explore the implications of this scenario in \S \ref{sec:varying}. $M_*$ can be related to the V-band luminosity using the mass function of UFDs \citep{Geha2013} via $M_*/L_V\approx 3$ (where both quantities are in solar units). This estimate applied to the sample of UFDs considered in this work results in $M_f\approx 0.1 M_{g,0}$.
	Therefore, we can expect to have $0.1 \lesssim M_g/M_{g,0} \leq 1$ at the time of the $r$-process event. We consider specific models for estimating $M_g/M_{g,0}$ in \S \ref{sec:gasloss}.

	We assume that the probability of an explosion occurring at a given location in the galaxy follows the stellar density distribution and apply a Monte Carlo method for calculating the fraction of material from the $r$-process event that remains in the galaxy after the explosion.
	For each realization of the Monte Carlo simulation, we draw an explosion location according to the stellar distribution and follow the blast-wave phases described in \S \ref{blastwavephases} to calculate the radius at which the explosion stops expanding. In order to avoid the complexities that would require a full hydrodynamical calculation we assume here that the density the explosion has to push through is spatially constant and equal to the density at the explosion center.
	We then calculate the volumetric fraction of the ``explosion sphere" that overlaps with $r_{90}$ which gives us a retainment fraction, $\xi$. This process is repeated $10^4$ times, to fully sample the distribution of explosion locations. We then calculate the mean value of $\xi$, which gives us the average retainment fraction for an explosion of energy $E$, occurring in a galaxy with $r_{1/2},\sigma_v$ and with a fraction $M_g/M_{g,0}$ of the initial gas remaining in the galaxy:
	\begin{equation}
	\xi\equiv \xi(\sigma_v,r_{1/2},E,\frac{M_g}{M_{g,0}})	
	\label{eq:xi}
	\end{equation}

\begin{figure*}
		\centering
		\includegraphics[scale=0.39]{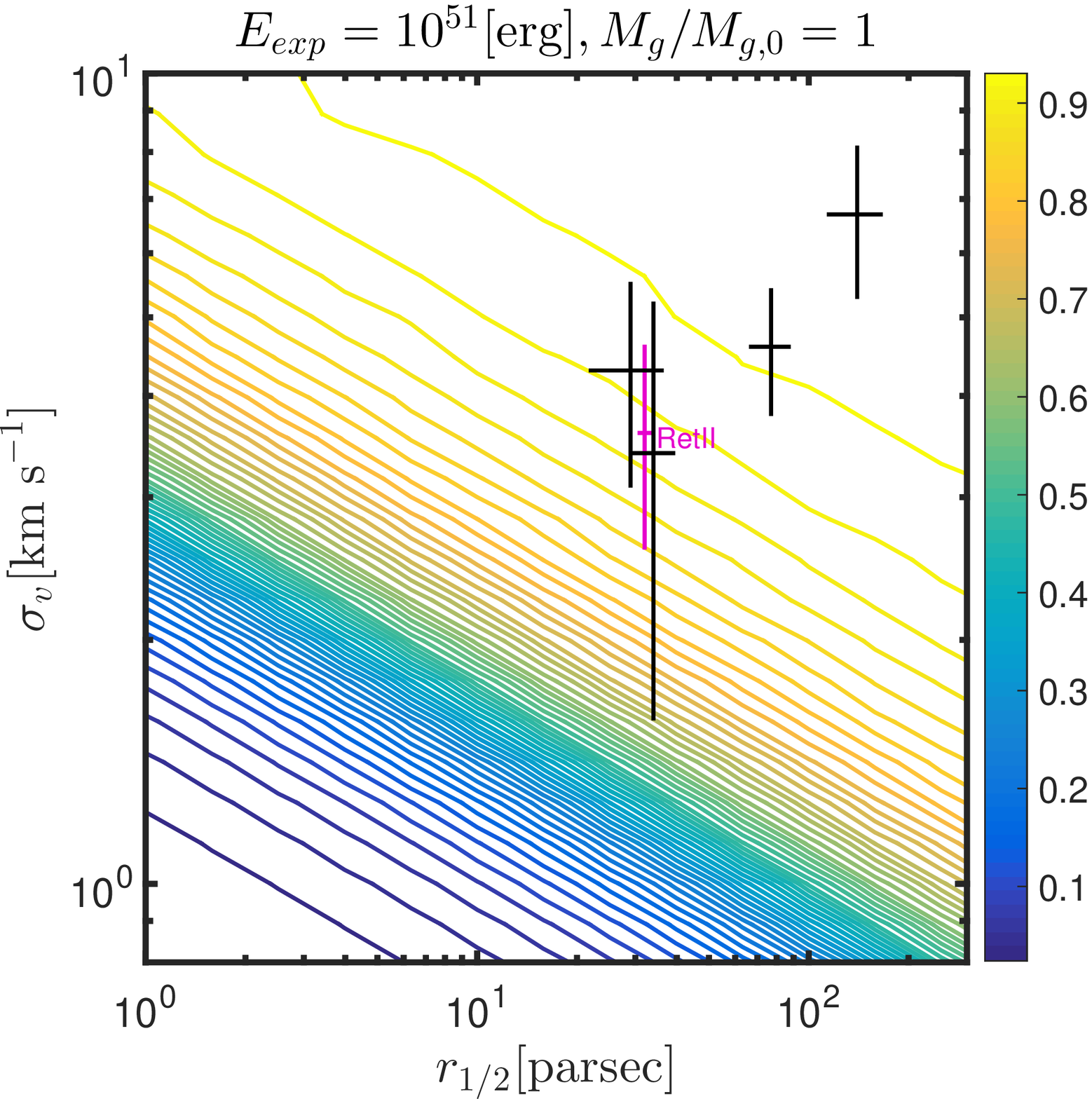}
		\includegraphics[scale=0.39]{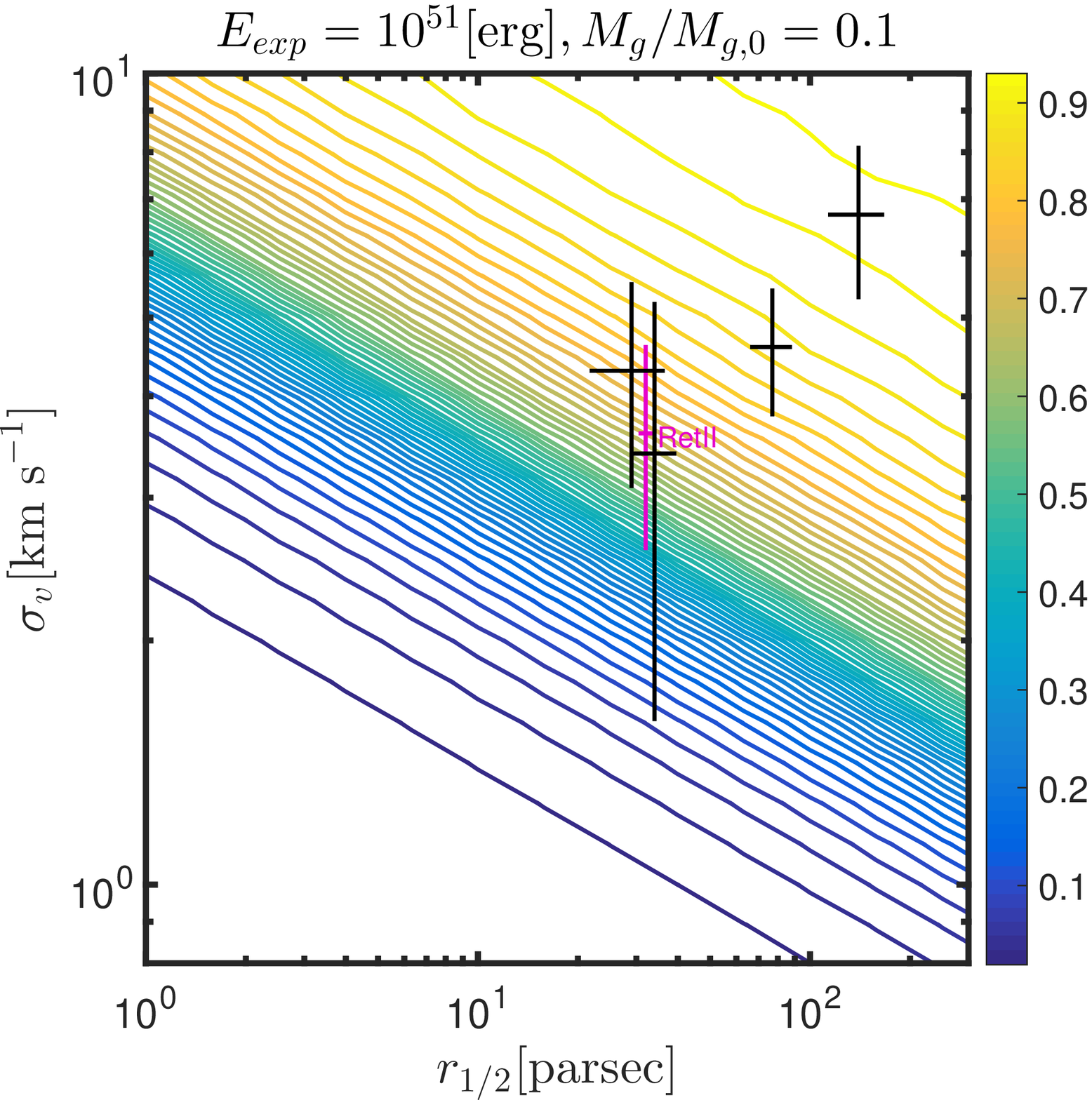}\\
		\includegraphics[scale=0.39]{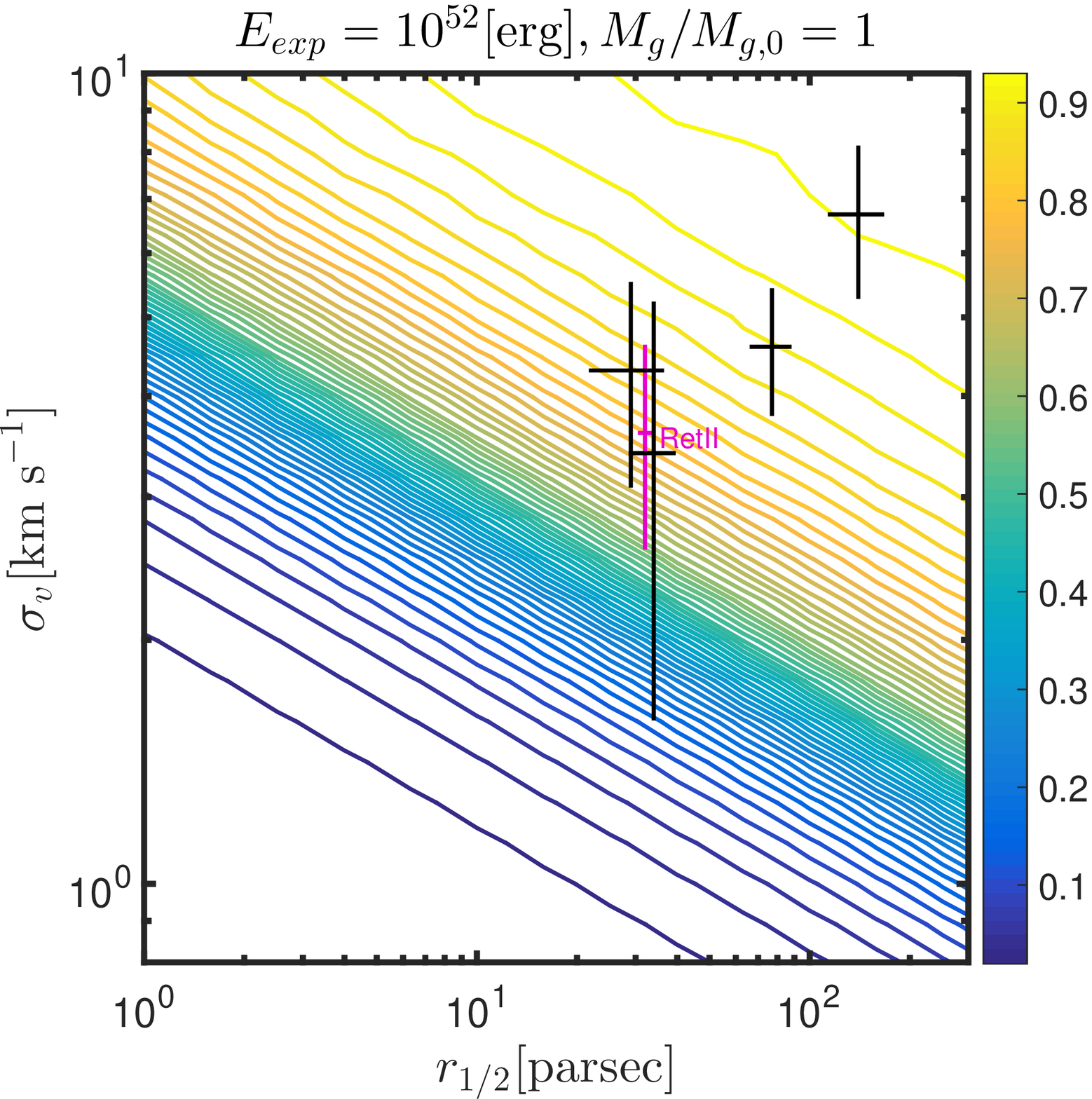}
		\includegraphics[scale=0.39]{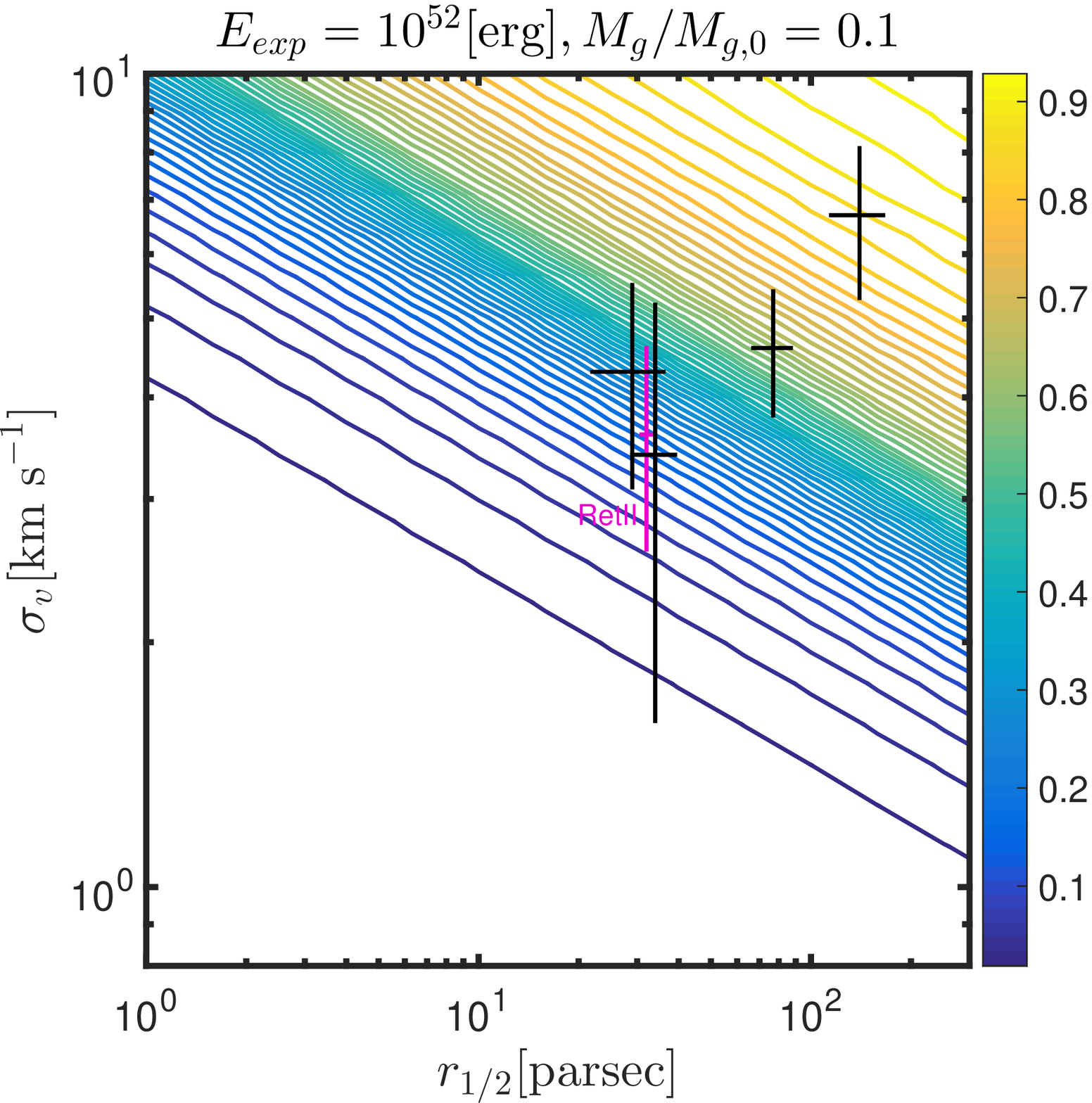}\\
		\includegraphics[scale=0.39]{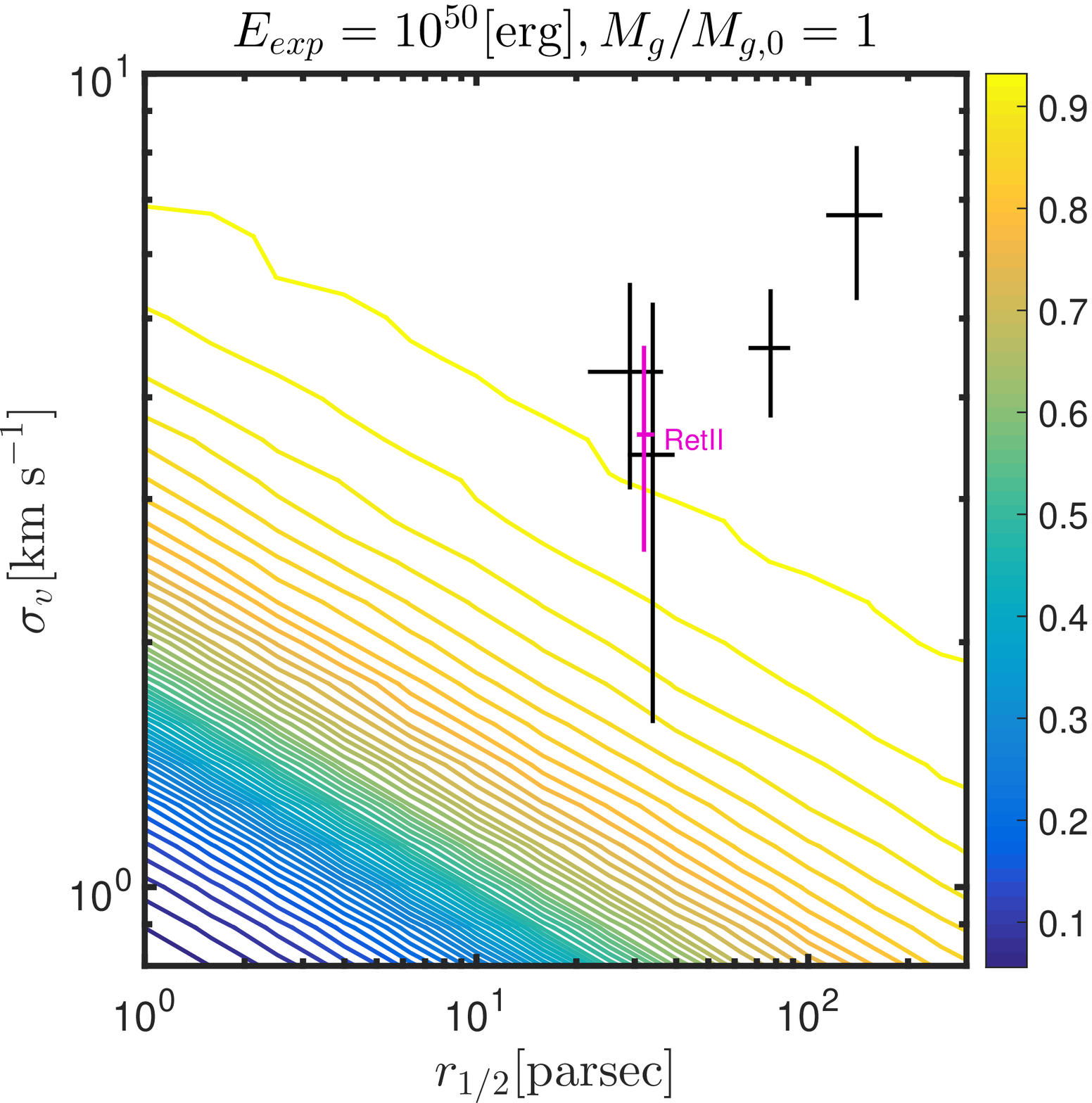}
		\includegraphics[scale=0.39]{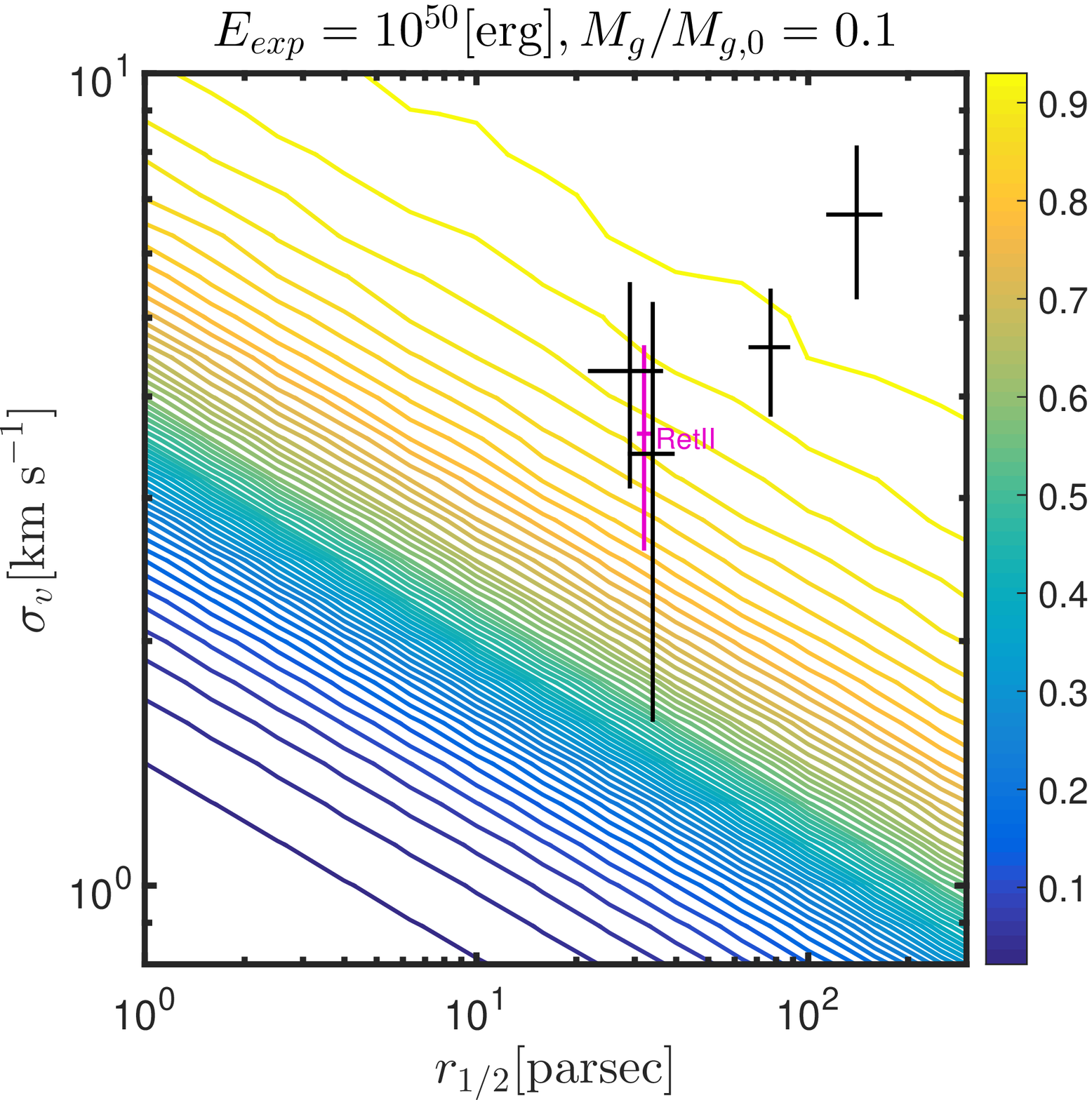}\\
		\caption
		{\small Fraction of $r$-process mass retained in galaxies with $\sigma_v,r_{1/2}$ for an $r$-process explosion energy $E_{\rm exp}$ and assuming that a fraction $M_g/M_{g,0}$ of the gas remains in the galaxy at the time of the $r$-process event. Black crosses denote observations of UFDs with no $r$-process material detected and Ret II is marked in purple.}
		\label{fig:retainment0}
	\end{figure*}

	Fig. \ref{fig:retainment0} depicts the retainment value, $\xi$, for different explosion energies and different vales of $M_g/M_{g,0}$. For an explosion with $E=10^{51}$erg and $M_g/M_{g,0}=1$ ($M_g/M_{g,0}=0.1$), the retainment value for Ret II is expected to be $\xi \approx 0.9$ ($\xi \approx 0.7$). For $E=10^{50}$erg the retainment fraction increases to $\xi \approx 0.93$ if $M_g/M_{g,0}=1$ (or $\xi \approx 0.86$ if $M_g/M_{g,0}=0.1$). Alternatively, for more energetic $r$-process events with $E=10^{52}$erg, $\xi \approx 0.77$ for $M_g/M_{g,0}=1$ (or $\xi \approx 0.16$ if $M_g/M_{g,0}=0.1$). As a comparison we consider also the retainment fraction for GW170817. As suggested by the discussion in \S \ref{blastwavephases}, the value for this considerably larger galaxy is now close to unity: $\xi \gtrsim 0.95$.

	It is useful to define an additional parameter,
	\begin{equation}
	\Psi\equiv\xi M_{g,0}/M_g\:.
	\end{equation}
	This parameter describes the ratio between the amount of mass implied by abundance observations in UFDs in two cases. The first, taking into account gas loss before the $r$-process event as well as retainment considerations, and the second, where both these considerations are absent (e.g., as in \cite{Beniamini2016}). $M_g<M_{g,0}$ will increase the observed abundance for a given synthesized mass, while $\xi<1$ means that only some of this $r$-process material really remains in the galaxy (thus working in the opposite direction, towards reducing the abundance). Interestingly, in all the cases considered above $\Psi>1$. Thus, the implied amount of $r$-process mass produced per event is reduced by $\Psi$ as compared with $6\times 10^{-3} M_{\odot}\lesssim \tilde{m}_{r-process}\lesssim 4\times 10^{-2} M_{\odot}$ obtained from UFD observations under the assumption of complete retainment and with $M_g=M_{g,0}$ \citep{Beniamini2016}. In order to keep the same amount of total $r$-process abundance (which is constrained by observations of classical dwarfs and the Milky Way), the rate of the events would have to be increased by the same amount. We return to the estimates of the synthesized mass and the rate in \S \ref{sec:likelihood} after exploring different models for mass loss in UFDs.

	\section{Modelling the gas loss in dwarf galaxies}
	\label{sec:gasloss}
\begin{figure*}
		\centering
		\includegraphics[scale=0.39]{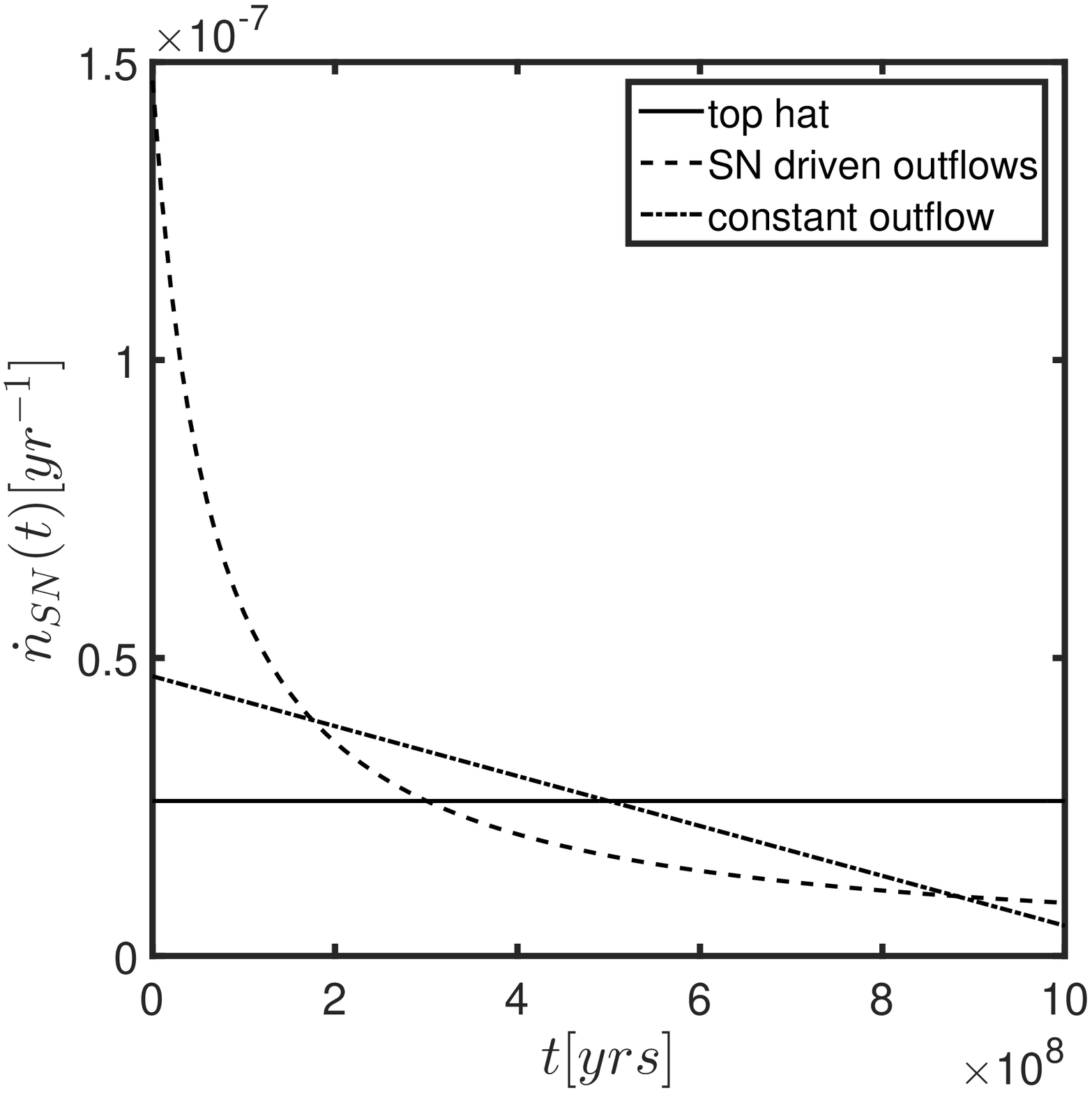}
		\includegraphics[scale=0.39]{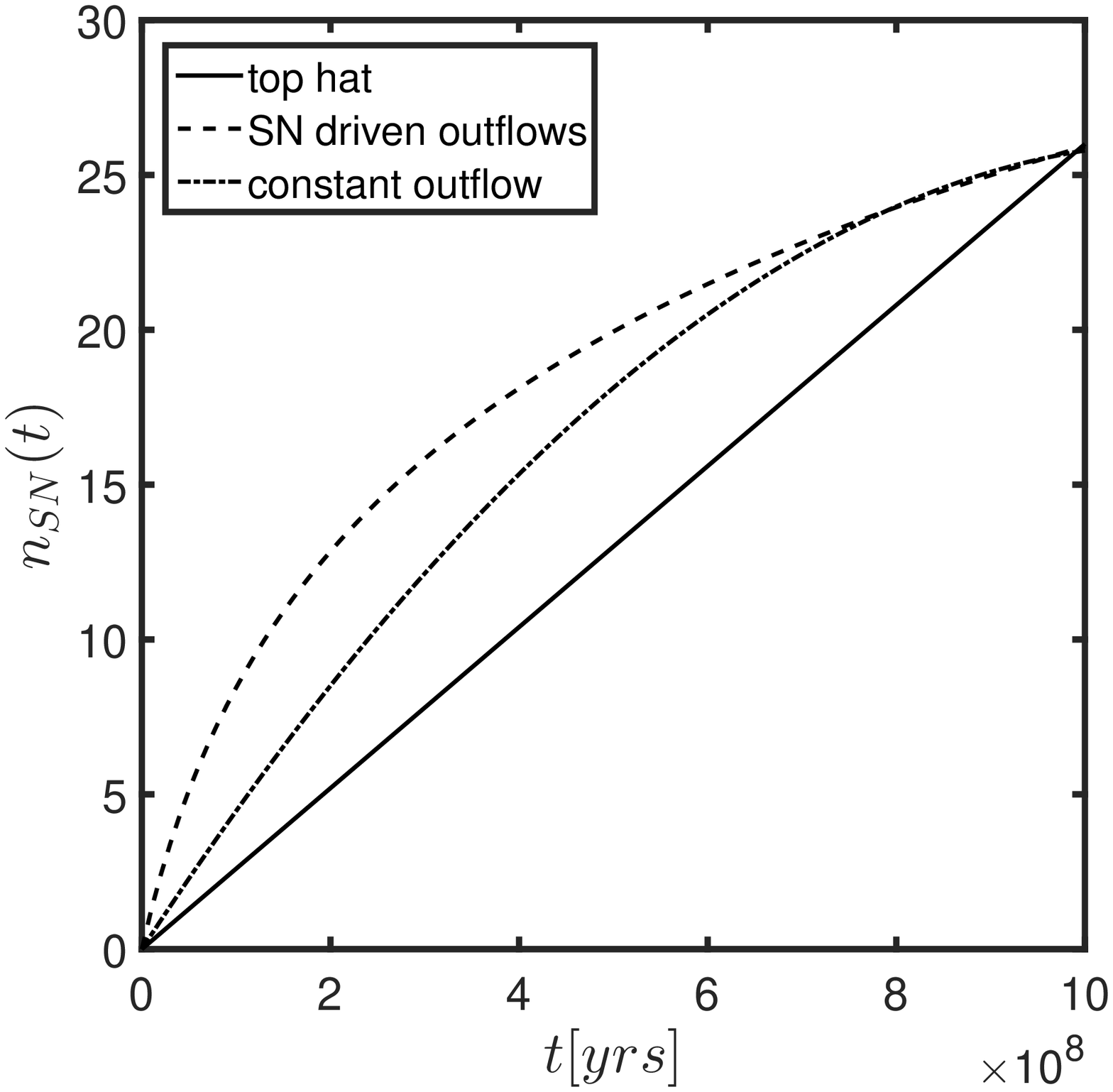}\\
		\caption
		{\small SN rate (left) and accumulated number of SNe for a Ret II-like galaxy and for the different outflow models discussed in this work.}
		\label{fig:nsn}
	\end{figure*}

	As shown above, the probability of retainment depends on the amount of gas that resides in the host dwarf galaxy at the time of explosion. We explore here 
	three different models for the mass loss in UFDs:
	\begin{enumerate}
		\item ``Top hat" evolution - The gas mass remains constant, then is completely removed on a short time-scale $\sim $Gyr after formation. 
		\item Removal of mass by SNe. Each SN is assumed to remove a constant fraction of the gas mass. Note that since we assume in what follows that the star formation rate is proportional  to the amount of the gas in the galaxy at a given time, this model implies that most of the star formation and SNe will occur early on. In particular, for our canonical model, $80\%$ of stars are produced within the first 500Myr and $90\%$ within 650Myr.
		\item A constant rate of mass outflow from the galaxy. 
	\end{enumerate}
	A comparison of these models is shown in Fig. \ref{fig:nsn}.
	In the following we describe these models in more detail.

	\subsection{Top hat evolution}
	\label{sec:tophat}
	The simplest model for mass loss is to assume that the gas mass remains constant at its initial level, and suddenly decreases over a brief time-scale towards the end of star formation. This model is less physically motivated than those discussed in \S \ref{sec:SNmassloss}, \S \ref{sec:constoutflow}. Its main purpose is to serve as a basis for comparison with previous works and with the other models of mass loss explored in this paper.
	The typical $r$-process to hydrogen abundance predicted by this model (assuming a single event occurs in a given galaxy and that full mixing occurs) is $n_{r/H}=m_r/M_{g,0}$  where $n_{r/H}$ is the relative density of some $r$-process element (e.g. Europium) as compared to hydrogen, and $m_r$ is the amount of that element's mass that is synthesised in one $r$-process event.
	
	\subsection{Mass loss by SNe}
	\label{sec:SNmassloss}
	The energy of a typical ccSNe \footnote{since star formation took place only in the first $\sim$ Gyr after the birth of these galaxies, type Ia SNe are unlikely to significantly contribute to the gas loss or iron production in these galaxies} is $\sim 10^{51}$erg. This is comparable or larger than the binding energy of UFDs which is approximately
	\begin{equation}
	U=-1.2G\frac{M_{1/2}^2}{r_{1/2}}\approx -3.4\times 10^{50} \bigg(\frac{\sigma_v}{3 \mbox{km s}^{-1}}\bigg)^4\bigg(\frac{r_{1/2}}{30 \mbox{pc}}\bigg)
	\end{equation}
	It is therefore reasonable to assume that a single SN can significantly disturb the gas in these galaxies and cause a certain fraction of that gas to become unbound. We adopt a simple model, whereby a constant fraction $f$ of the galaxy's gas remains after each ccSN.
	We assume that the iron production of ccSNe is similar to that measured by SNe observations in the local universe, i.e. 2/3 of SNe are type II and produce $0.02M_{\odot}$ \citep{Drout2011,Kushnir2015} and 1/3 are type Ib,Ic and produce $0.2M_{\odot}$ \citep{Li2011}. We further assume that, as implied by observations of double NS (DNS) systems in the galaxy, $60-70\%$ of stellar collapses to neutron stars are ultra-stripped SNe that involve virtually no iron production \citep{BP2016}. Overall, the mean amount of iron per ccSNe is approximately $m_{Fe}=0.03M_{\odot}$.

	The value of $f$ for each UFD galaxy can be estimated from the total V-band stellar luminosity of UFD galaxies, $L_V$. As described in \S \ref{sec:retainmentfrac} we assume that the gas mass required to create the stars in UFDs is roughly ten times larger than the resulting stellar mass observed in those galaxies (notice that the final value of $f$ depends only logarithmically on this number). This implies that
	\begin{equation}
	\label{eq:whatisf}
	\frac{M_f}{M_{g,0}}=10\frac{M_*}{L_V}\frac{L_V}{M_{g,0}}=f^{N_{SN}}
	\end{equation}
	where  $M_*/L_V=3$ in solar units (see \S \ref{sec:retainmentfrac}) and $N_{SN}$ can be related to the V-band luminosity of dwarf galaxies using the initial mass function (IMF) of UFDs \citep{Geha2013} and is approximately $N_{SN}\approx 0.026 L_V$. Using Eq. \ref{eq:whatisf} we can now estimate $f$ from the observed parameters.
	Under the assumption that gas loss is done predominantly through SNe and that $f$ is constant throughout the galaxy's evolution, we find that for Ret II $f\approx 0.92$ and for the entire sample of UFDs considered here $\langle f \rangle =0.9$. Thus, observations imply that about $10\%$ of the gas is expelled by each SN occurring in these galaxies.

	We can now calculate the relative iron to hydrogen density ratio, $n_{Fe/H}$ after $N$ SNe using
	\begin{eqnarray}
	\label{eq:nFeH}
	& n_{Fe/H}=\frac{M_{Fe}}{M_{g}}=\frac{m_{Fe}\sum_{i=1}^{n}f^{n+1-i}\xi(\sigma_v,r_{1/2},E,f^{i-1})}{M_{g,0} f^{n}} \nonumber \\
	& =\frac{m_{Fe}}{M_{g,0}}\sum_{i=1}^{n} f^{1-i}\xi(\sigma_v,r_{1/2},E,f^{i-1})
	\end{eqnarray}
	where $\xi$ is the average retainment fraction (see Eq. \ref{eq:xi}). Note that $n$ here can be anywhere in the range $0\leq n\leq N_{SN}$, depending on how many SNe exploded before the formation of any given star. Notice that the advantage of writing the abundance in terms of $n$ (rather than explicitly as a function of time), is that if the SNe rate is proportional to the star formation rate (as will be assumed below), then there is a one to one mapping between the number of stars with given abundances and the number of SNe that preceded them (e.g. half the stars are born by $n=N_{SN}/2$ and so on).
	The further back in this process a SN occurred, the more times the iron that was synthesized in that SN was partially expelled from the galaxy. At the same time, the more mass has been removed from the galaxy, the harder it is to retain Fe from future SNe. Since the former effect is stronger than the latter (see \S \ref{sec:retainmentfrac}), the last SN preceding the birth of any given star contributes the most towards its abundance.
	For a given observed value of $n_{Fe/H}$, the required $n$ increases with $f$ and  $n \leq n(f=1)=M_{g,0} n_{Fe/H}/(m_{Fe}\xi)$. Stated differently, for a given $n$, $n_{Fe/H}$ decreases with $f$.

	We turn now to estimate the abundance of $r$-process elements. Given that $r$-process events are much rarer than ccSNe, with the relative rate being $2.5\times 10^{-4}\lesssim R_{rp/SN}\lesssim 1.4\times 10^{-3} $ \citep{Beniamini2016}, we can estimate the probability of $k$ $r$-process events taking place before $1\leq n\leq N_{SN}$ SNe exploded, using Poisson statistics,
	\begin{equation}
	\label{eq:Poisson}
	P_r(k,n)=e^{-nR_{rp/SN}}\frac{(nR_{rp/SN})^k}{k!}.
	\end{equation}
	Since for any of the known UFDs, the observed iron abundances imply $N_{SN}\lesssim 100$, we have that generally $nR_{rp/SN}<N_{SN} R_{rp/SN} \ll 1$. We therefore expect that at most one $r$-process event takes place in such galaxies and that it typically occurred after $\sim N_{SN}/2$ SNe have exploded. {We note that by definition for the case of NS-NS mergers, at least two SNe took place before the $r$-process event. In this case, equation \ref{eq:Poisson} can be applied under the assumption that $n$ is the number of independent SNe, i.e. assuming the true number is $m=n+2$. For the canonical case of $\langle f \rangle=0.9$, this introduces only a minor correction. It will become more important for smaller values of $f$, as explored in \S \ref{sec:varying}.

	This allows us to calculate the relative increase of $r$-process abundance as compared with the case of no mass loss and perfect retainment
	\begin{equation}
	\label{eq:Psi}
	\Psi\approx\xi(\sigma_v, r_{1/2},E,f^{N_{SN}/2})f^{-N_{SN}/2}\:.
	\end{equation}
	For Ret II, the result is $\xi \approx 0.84$ and $\Psi \approx 2.5$.

	We turn now to estimate the average expected abundance of $r$-process elements. We denote the supernovae rate (number of SNe per unit time) by $\dot{n}_{SN}(t)$ (and equivalently $n_{SN}(t)$ is the number of SNe that exploded up to a time $t$). For simplicity, we assume here that $\dot{n}_{SN}(t)$ is proportional to the star formation rate, which is constant per unit mass up to a time $T_*$, after which it drops to zero. This implies $\dot{n}_{SN}(t)\propto M_g(t)$ and using $M_g=M_{g,0}f^{n_{SN}(t)}$ we obtain a differential equation for $n_{SN}(t)$ 
	\begin{equation}
	\dot{n}_{SN}(t) = A(f) f^{n_{SN}(t)}
	\end{equation}
	where $A(f)$ is a normalization constant that can be determined by the total number of SNe that occurred in the UFD (which in turn can be gauged by its total stellar luminosity). Assuming $f<1$, this equation has the solution (see also Fig. \ref{fig:nsn})
	\begin{equation}
	\label{eq:diffsol}
	1-f^{-n_{SN}(t)}=A(f) t \log(f)
	\end{equation}
	which at $T_*$ gives $ 1-f^{-N_{SN}(t)}=A(f) T_* \log(f)$,
	where $N_{SN}=n_{SN}(T_*)$ is the total number of SNe.
	Denoting by $r_0$ the SNe rate implied by the total amount of stars formed when $f=1$ (i.e. $r_0=N_{SN}(f=1)/T_*\approx 0.026L_V/T_*$) and imposing that the same amount of stars are formed for $f<1$ ($\int \dot{n}_{SN}(t)dt=r_0 T_*$) we find $A(f)$:
	\begin{equation}
	A(f)=\frac{1-f^{-r_0T_*}}{T_*\log(f)}.
	\end{equation}

	Assuming that a single $r$-process event takes place in a UFD (and that on average it happens after $N_{SN}/2$ SNe) we obtain
	\begin{equation}
	\label{eq:nrH}
	n_{r/H}=\frac{m_{r}}{M_{g,0}f^{N_{SN}/2}}\xi=\frac{m_{Eu}}{M_{g,0}}\xi f^{-r_0T_*/2},
	\end{equation}
	Notice that $\xi$ increases with $f$ while at the same time $f^{-r_0 T_*/2}$ decreases. Since the latter is the dominating factor, the relative $r$-process abundance is a decreasing function of $f$.
	
	\subsection{Constant mass loss rate}
	\label{sec:constoutflow}
	We consider here the possibility of a constant outflow of gas out of the galaxy 
	\begin{equation}
	M_g(t)=M_{g,0}-\frac{M_{g,0}-M_f}{T_*}t
	\end{equation} 
	As in \S \ref{sec:gasloss} we assume that $\dot{n}_{SN}(t)\propto M_g(t)$. This implies that 
	\begin{equation}
	n_{SN}(t)=\frac{N_{SN} (M_{g,0}T_*t-\frac{1}{2}(M_{g,0}-M_f)t^2)}{\frac{1}{2}T_*^2(M_{g,0}+M_f)}
	\end{equation}
	see also Fig. \ref{fig:nsn}.
	Assuming once more that an $r$-process event takes place on average after $N_{SN}/2$ SNe have occurred, the mean time of the event and the gas mass in the galaxy at that time are
	\begin{eqnarray}
	& \langle t_{rp} \rangle =\frac{T_*}{M_{g,0}-M_f}\bigg(M_{g,0}-\sqrt{\frac{1}{2}(M_{g,0}^2-M_f^2)}\bigg) \\
	& M_g(\langle t_{rp} \rangle)=\sqrt{\frac{1}{2}(M_{g,0}^2-M_f^2)}
	\end{eqnarray}
	For Ret II, this leads to $M_g/M_{g,0}=0.7$ and $\Psi=\xi M_{g,0}/M_g\approx 1.3$. Finally, the relative $r$-process to hydrogen abundance is
	\begin{equation}
	n_{r/H}=\frac{m_r}{\sqrt{\frac{1}{2}(M_{g,0}^2-M_f^2)}} \xi
	\end{equation}
	which is slightly larger than for the top hat case.

	\section{$r$-process event rates and mass per event}
	\label{sec:likelihood}
	We apply here a likelihood analysis in order to estimate the best fit parameters for the $r$-process event rate and the $r$-process mass created per event. We use Eu as a representative $r$-process element, as it is known to be predominantly synthesized via the $r$-process and there are reliable measurements and/or upper limits for its abundance. We consider the same sample of UFDs with observational constraints of $\sigma_v$, $r_{1/2}$, $\mbox{[Fe/H]}$, $\mbox{[Eu/H]}$ and $L_V$ considered by \cite{Beniamini2016} (see description of the sample in \S \ref{sec:retainmentfrac}). This sample includes Ret II in which $r$-process elements were detected and four other UFDs for which there are upper limits on [Eu/H]. We define a likelihood function to be maximized
	\begin{equation}
	L(R_{rp/SN},m_{Eu},f)=\Pi_i P_i(n_{Eu/H},n_{Fe/H}|R_{rp/SN},m_{Eu})
	\end{equation}
	where the product runs over the UFD galaxies in our sample, and $P_i$ is the probability of obtaining the observed Fe and Eu abundances in the $i-$th galaxy for given values of the model parameters: $R_{rp/SN},m_{Eu}$.

	To calculate $p_i$ we apply a Monte Carlo method. For each galaxy we estimate the SNe rate from $L_V$ (see \S \ref{sec:gasloss}) and extract the $r$-process event rate using the relative $r$-process to SNe rate, $R_{rp/SN}$. We use Poisson statistics to calculate the probabilities of obtaining a given number of events from those rates and the times at which they occur. Finally, integrating over the star formation rate between each two events, allows us to estimate the number of stars with given Fe, Eu abundances and consequently we obtain their mean values. We repeat this process $10^5$ times for each set of model parameters and measure the fraction of realizations which result within the observed values of [Fe/H],[Eu/H]. This provides us with a value for $P_i$.

	The results of the likelihood analysis are shown in Fig. \ref{fig:likelihood} for the three different mass loss models considered in this work.
	The $2\sigma$ limits on the best fit parameters are:
	\begin{enumerate}
		\item Top hat model -  $2\times 10^{-4}M_{\odot}\lesssim R_{rp/SN}\lesssim 1.5\times 10^{-2}$, $5\times 10^{-6}\lesssim m_{Eu}\lesssim 4\times 10^{-5} M_{\odot}$. This model is similar to the one considered by \cite{Beniamini2016}, with the main change being due to adding the possibility of $\xi<1$ (i.e. not all Fe or Eu are retained in the galaxy). The results reported here are consistent with the values found in that work.
		\item SN driven outflows - $2\times 10^{-4}M_{\odot}\lesssim R_{rp/SN}\lesssim 1.5\times 10^{-2}$, $10^{-6}\lesssim m_{Eu}\lesssim 2\times 10^{-5} M_{\odot}$. The main difference from the previous case, is a value of $m_{Eu}$ lower by $\sim 2$. This is due to the typical value of $\Psi=\xi M_{g,0}/M_g\approx 2.5$ at the time of the $r$-process event (see \S \ref{sec:SNmassloss}).
		\item Constant outflow - $2\times 10^{-4}M_{\odot}\lesssim R_{rp/SN}\lesssim 1.5\times 10^{-2}$,  $2\times 10^{-6}\lesssim m_{Eu}\lesssim 3\times 10^{-5} M_{\odot}$. This model results in values that are somewhat closer to the top hat model. Once more, this can be related to the value of $\Psi\approx 1.3$ at the time of the event, which is closer now to unity.
	\end{enumerate}

	All three models provide similar results and are statistically consistent with the available data. We also compare the preferred values of the rate and Eu mass per event with the limits from the recently observed binary NS merger, GW170817 \citep{2017arXiv171005443D,2017arXiv171005858P}. Notice that the luminosity and  time-scale of the observed macronova constrain the total amount of $r$-process materials created in the merger, and the opacity of the ejecta, that may become dominated by lanthanide elements with $A>140$. As opposed to the abundance measurements in UFDs, these observations do not directly constrain the Eu mass. Thus in order to convert the former to an Eu mass, we need to assume the minimum atomic number of the $r$-process elements that are synthesized in the merger. The main uncertainty is whether or not the so called first peak $r$-process elements are created. We therefore assume either $A_{\rm min}=69$ or $A_{\rm min}=90$ to calculate $m_{Eu}$.
	We turn now to estimating the rate relative to SNe. The observed neutron star merger rate as implied by GW170817 is $r_{\rm NSM,obs}=1540_{-1220}^{+3200}\mbox{Gpc}^{-3}\mbox{yr}^{-1}$. Note, however, that the upper bounds on the rate may be somewhat optimistic, given the limits from the Swope Supernova survey 1a for optical counterparts similar to that of GW170817 \citep{Siebert2017}, which suggest $r_{\rm NSM,obs}<1.6\times 10^4\mbox{Gpc}^{-3}\mbox{yr}^{-1}$. We use the local observed core-collapse SNe rate of $r_{\rm ccSNe,obs}=7.1^{+1.4}_{-1.3}10^4 \mbox{Gpc}^{-3}\mbox{yr}^{-1}$ \citep{Drout2011,Li2011}. For consistency with our assumptions in \S \ref{sec:gasloss}, this rate needs to be converted to the intrinsic collapse rate, assuming $f_{\rm weak}=0.6-0.7$ of stellar collapses are ultra-stripped SNe \citep{BP2016} and thus unobservable. This is also in accord with the 'missing SNe problem' \citep{Horiuchi2011ApJ}. Finally, only a certain fraction $f_{\rm Merg}$ of the general population of merging double neutron stars are expected to both remain confined in UFD galaxies given their natal kicks, and merge before the end of star formation in their host galaxy. \cite{Beniamini2016a} estimate this fraction to be $0.06\lesssim f_{\rm merge}\lesssim 0.6$. Putting it all together we find
	\begin{equation}
	R_{rp/SN}=f_{\rm Merg}f_{\rm weak}\frac{r_{\rm NSM,obs}}{r_{\rm ccSNe,obs}}\approx 0.0028_{-0.019}^{+0.0025}
	\end{equation}
	where we have used $f_{merge}=0.2$ as the canonical value \citep[see also][for a discussion on r-process event rates in a cosmological setting and implications for GW observations]{2016MNRAS.455...17V}.
	As can be seen in Fig. \ref{fig:likelihood}, our results for the UFD sample are marginally consistent with the values inferred from GW170817 for $A_{\rm min}=69$ and result in somewhat lower masses per event than the values inferred for $A_{\rm min}=90$. We note however, that as shown by \cite{Beniamini2016}, including classical dwarfs together with ultra faint dwarfs increases the estimates for the mass per event by a factor of $\sim 6$, while keeping the rate fixed. Taking this effect into account will result in consistency with observations of GW170817 assuming either $A_{\rm min}=69$ or $A_{\rm min}=90$. Furthermore, given that at the time of writing there is only one UFD with observed $r$-process abundance and only one clear determination of $r$-process heated macronova in association with a GW event, it remains undetermined whether or not there is any true discrepancy here.
	
	\begin{figure*}
		\centering
		\includegraphics[scale=0.39]{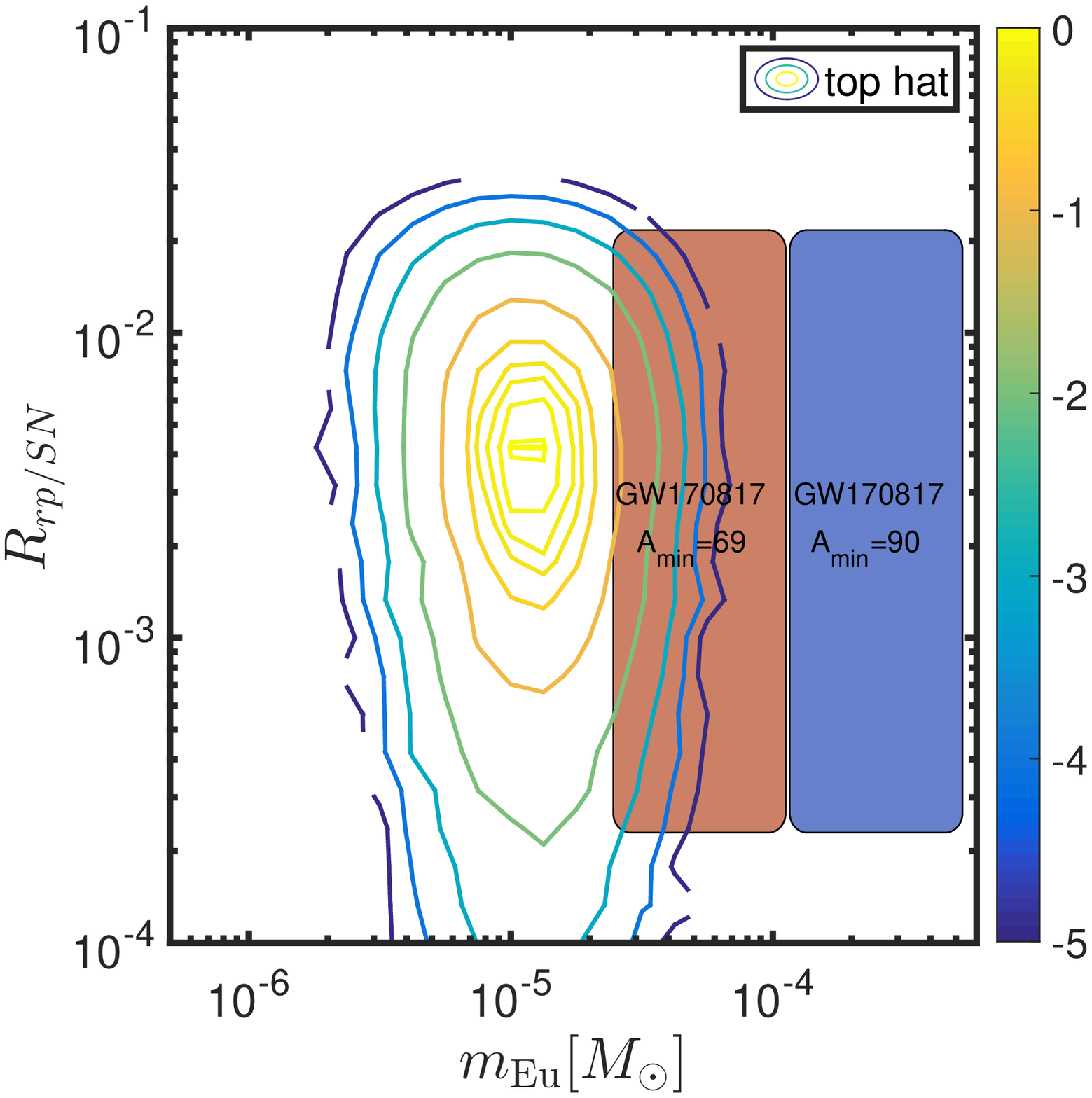}
		\includegraphics[scale=0.39]{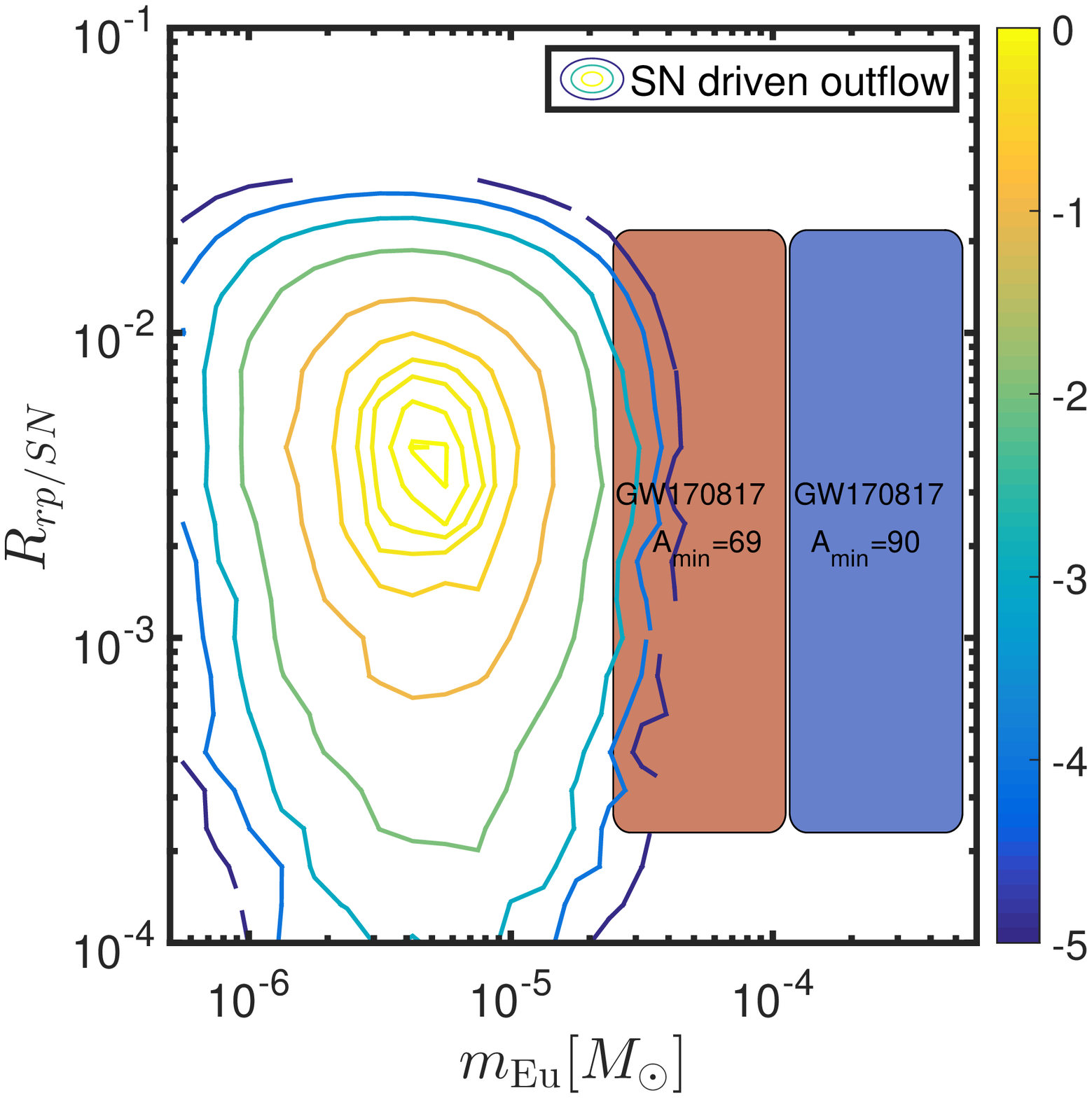}\\
		\includegraphics[scale=0.39]{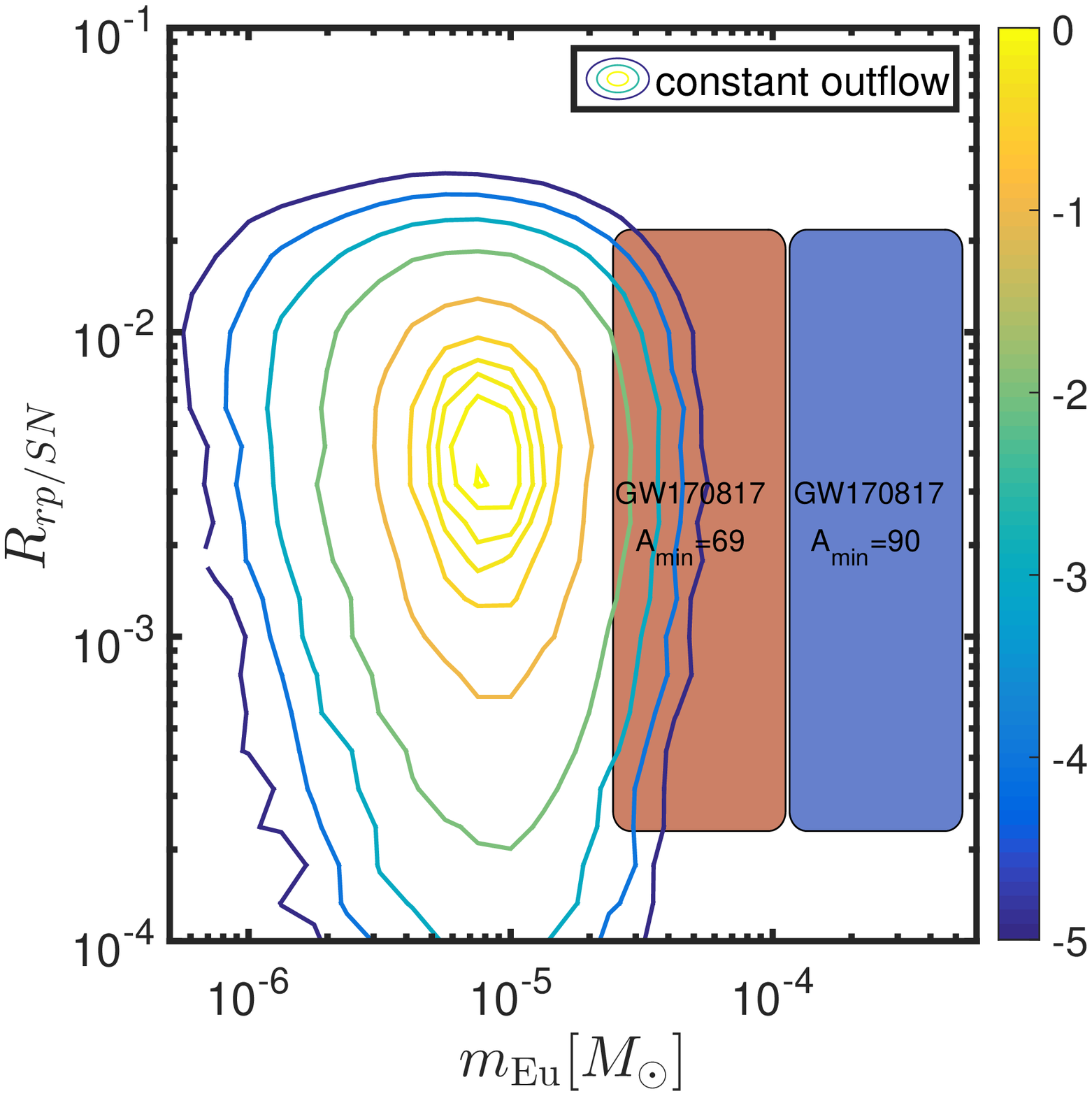}
		\caption
		{\small Log-likelihood  for $r$-process production as a function of the Eu mass per event $m_{Eu}$ and the relative $r$-process to ccSN event rate, $R_{rp/SN}$, for UFD galaxies and the different models of mass loss explored in this work. Also depicted are the inferred rate from GW170817.}
		\label{fig:likelihood}
	\end{figure*}

	\section{How rare is Ret II?}
	\label{sec:rareRetII}
	Given that Ret II is the only UFD galaxy with measured values of $r$-process materials, and that other UFDs, some of which are more luminous than Ret II, only have upper limits on their $r$-process abundances, it is clear that Ret II - like objects are rare. We set here to explore exactly how rare should such galaxies be.

	We perform a Monte Carlo simulation in which we use the observed $\sigma_v,r_{1/2}$ of Ret II and use a Poisson distribution for the $r$-process event rate and a log-normal distribution for the $r$-process mass per event with
	the mean values informed by the likelihood analysis in \S \ref{sec:likelihood} (and with the width of the log-normal distribution equal half the mean). We then draw $10^5$ realizations and calculate the cumulative distribution functions of Fe and Eu. Fig. \ref{fig:FeHEuH} depicts the results of these distributions compared with the observed values of [Fe/H], [Eu/H] in Ret II.
	As expected from Eq. \ref{eq:nFeH}, the top hat model typically produces lower iron abundances than the other two models. The expected (2$\sigma$) ranges of [Fe/H] are $-3.1\lesssim \rm{[Fe/H]}\lesssim -2.5$ for the top hat model, $-2.9\lesssim \rm{[Fe/H]}\lesssim -2.4$ for the SN driven outflows model and $-2.7\lesssim \rm{[Fe/H]}\lesssim -2.4$ for the constant outflow model. For the top hat model, the fraction of realizations resulting in a value of [Fe/H] consistent with the $1\sigma$ limits from observations is $0.93$. This fraction reduces to $0.51$ ($0.27$) for the SN driven outflows (constant outflow) model. 
	In all three models, the probability of having an $r$-process event in Ret II is $\sim 7.5\%$. Out of these, $37\%-41\%$ of models result in values of [Eu/H] consistent with the $1\sigma$ observational limits for Ret II.
	The expected (2$\sigma$) ranges of [Eu/H] in case there is at least one $r$-process event are $-2.2\lesssim \rm{[Eu/H]}\lesssim -0.85$ for the top hat model, $-1.83\lesssim \rm{[Eu/H]}\lesssim -0.79$ for the SN driven outflows model and $-1.66\lesssim \rm{[Eu/H]}\lesssim -0.85$ for the constant outflow model.
	The overall probability of obtaining abundances of both Fe and Eu within the $1\sigma$ observational errors is between $8\times 10^{-3}$ and $2\times 10^{-2}$ in all three models.

	\begin{figure*}
		\centering
		\includegraphics[scale=0.39]{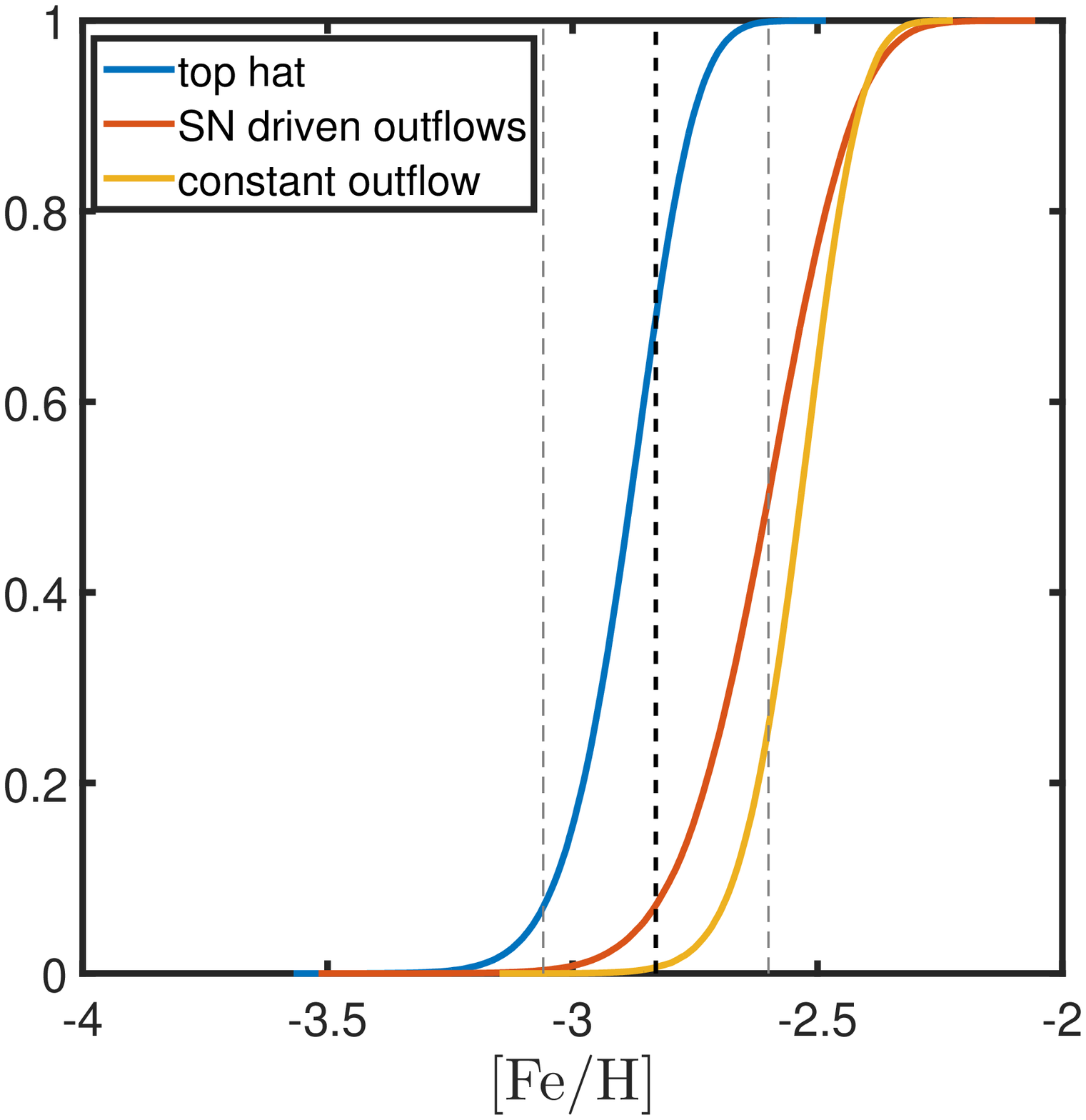}
		\includegraphics[scale=0.39]{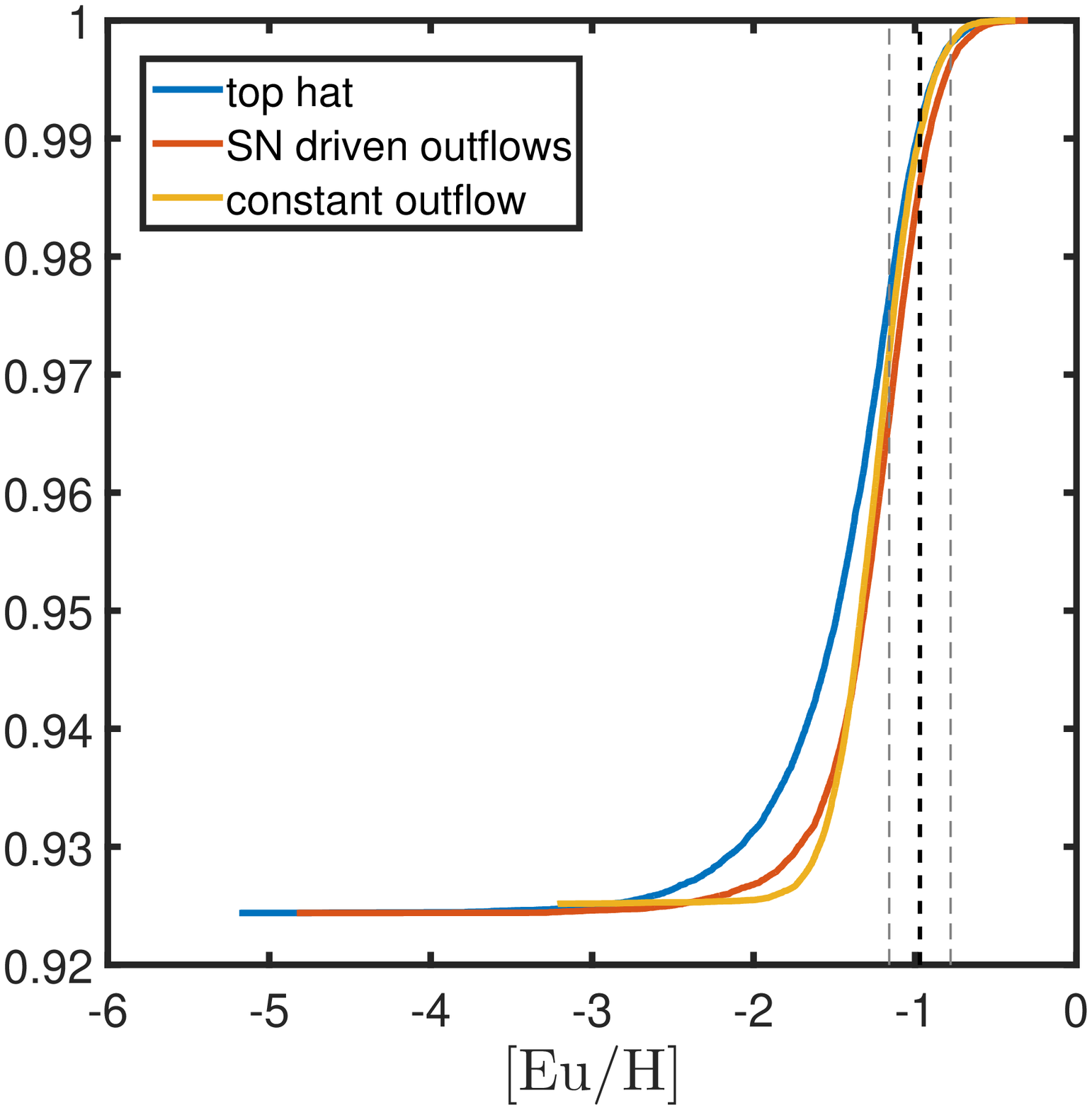}\\
		\caption
		{\small Cumulative distribution of [Fe/H] and [Eu/H] in an ensemble of galaxies with the same observed properties as Ret II and using the best fit values for the Eu mass per event and $r$-process rate as given by the likelihood analysis in \S \ref{sec:likelihood}. Dashed black lines mark the mean measured abundance for Ret II and dashed gray lines depict $1\sigma$ limits on these values.}
		\label{fig:FeHEuH}
	\end{figure*}
	
	\section{Varying model parameters}
	\label{sec:varying}
	We explore here the sensitivity of the results found above
	on the assumed parameters of the model.

	An important parameter in the model involving SNe driven outflows is $f$, the fraction of gas retained in the galaxy after each SNe. 
	As mentioned in \S \ref{sec:retainmentfrac}, if all the gas of UFDs is re-processed into stars within a duration $\ll 1$Gyr (see however \citealt{Brown2014,Weisz2015}), the fraction of gas removed by SNe can become much larger, and the implied value of $f$ becomes smaller. At this limit, $M_f$ should represent the observed gas in UFDs at the present age. Using equation \ref{eq:whatisf} with $M_f=10M_{\odot}$ \citep{Spekkens2014}, we find that $\langle f \rangle=0.7$ in this case. We re-do the likelihood analysis presented in \S \ref{sec:likelihood} for this scenario. The general trend is clear. Reducing $f$ implies that more gas is removed from the galaxies at earlier stages. As a result, the $r$-process event takes place when $M_g\ll M_{g,0}$, leading to less retainment of heavy elements, but more importantly, to mixing of those materials within a smaller amount of mass (i.e. $\Psi>1$, see discussion in \S \ref{sec:retainmentfrac}). Thus, the required $m_{Eu}$ is reduced as compared with the canonical case.
	The iron abundance is significantly increased (by a factor of $\sim 3$) in this scenario, due to the fact that the iron is mixed with a smaller amount of gas before that gas is re-processed into stars. The 2$\sigma$ limits of the likelihood analysis with $\langle f \rangle=0.7$ in terms of the rate and mass per event are  $1.5\times 10^{-3}M_{\odot}\lesssim R_{rp/SN}\lesssim 3.5\times 10^{-2}$, $2\times 10^{-7}\lesssim m_{Eu}\lesssim 5\times 10^{-6} M_{\odot}$. These results are in contention with the results from the observations of GW170817 and with constraints from the MW and classical dwarf galaxies on the total amount of $r$-process mass $\propto R_{rp/SN}\times m_{Eu}$ (see \cite{Hotokezaka2018} and references within). Furthermore, this model provides a poor description of the available data as compared with the canonical data, and can be ruled out statistically at a confidence level of $95\%$. This can be seen visually in Fig. \ref{fig:likelihoodf07}, which demonstrates that the (2$\sigma$) ranges of [Fe/H] and [Eu/H] for a Ret II like galaxy, with the best-fit parameters of the $\langle f \rangle=0.7$ model are: $-2.6\lesssim \rm{[Fe/H]}\lesssim -2.1$, $-3.3\lesssim \rm{[Eu/H]}\lesssim -0.9$ in strong contention with the observed values.
	
	\begin{figure*}
		\centering
		\includegraphics[scale=0.39]{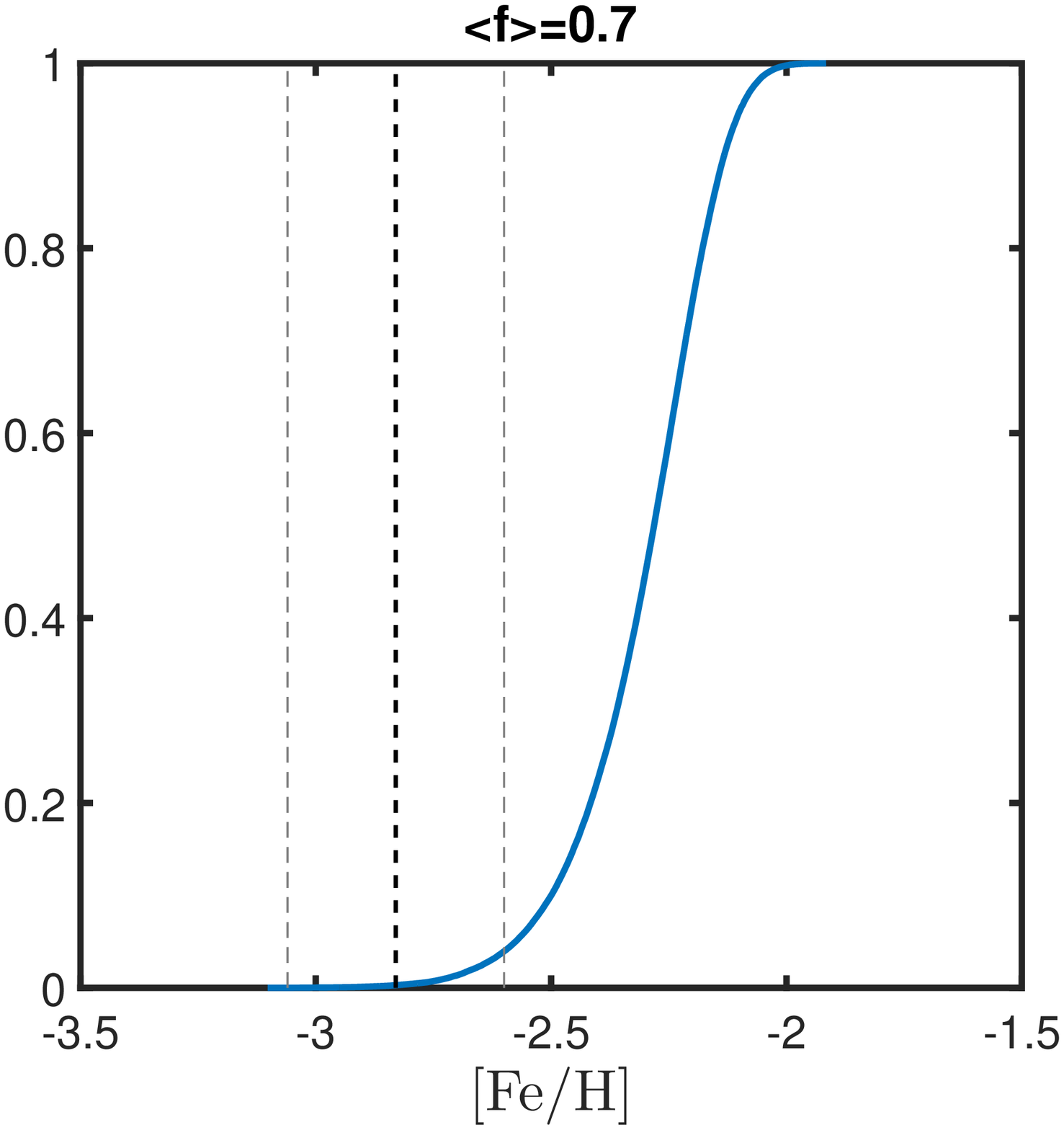}
		\includegraphics[scale=0.39]{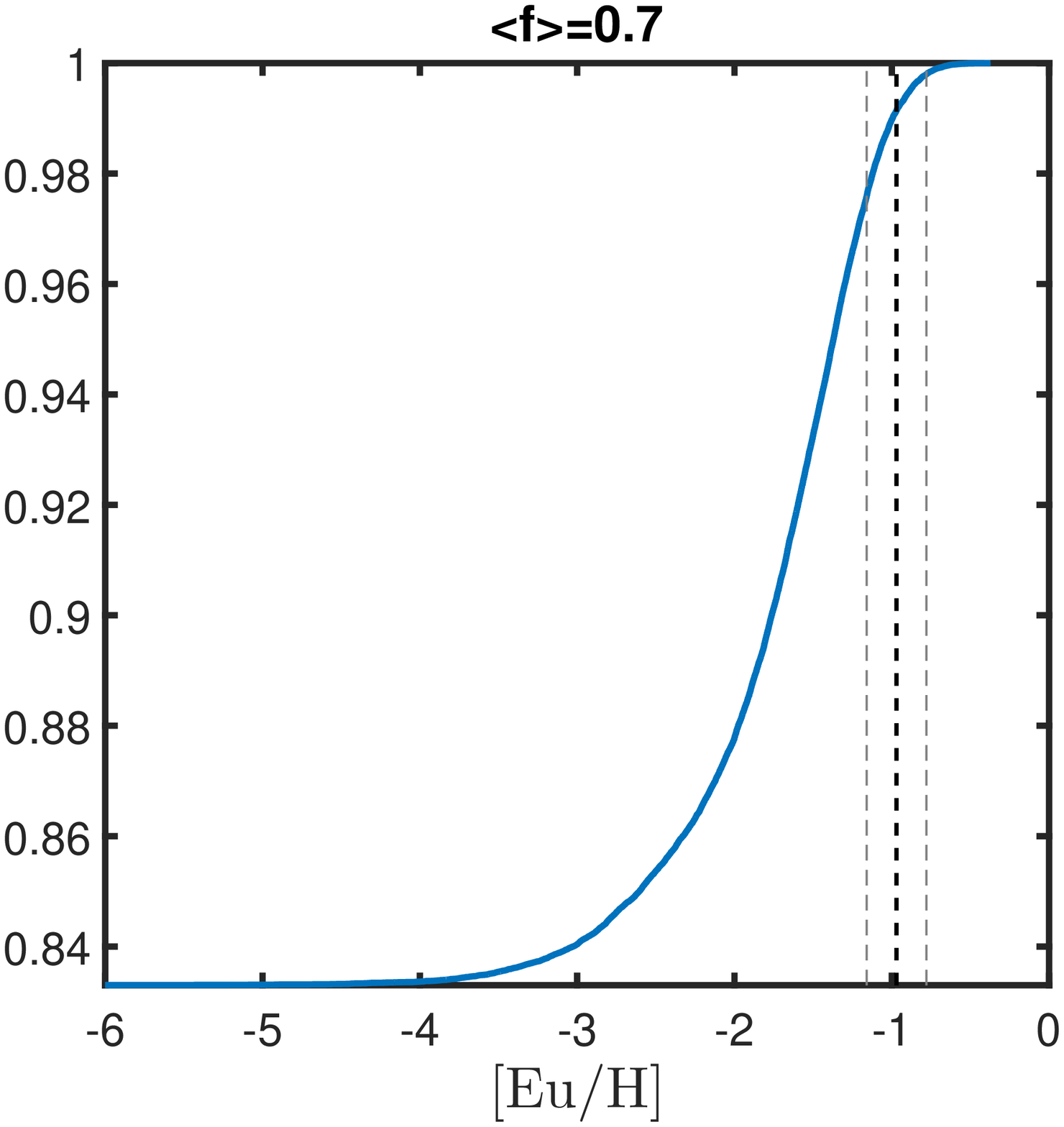}
		\caption
		{\small Cumulative distribution of [Fe/H] and [Eu/H] in an ensemble of galaxies with the same observed properties as Ret II and using the best fit values for the Eu mass per event and $r$-process rate assuming SNe driven outflows and $\langle f \rangle=0.7$. Dashed black lines mark the mean measured abundance for Ret II and dashed gray lines depict $1\sigma$ limits on these values.}
		\label{fig:likelihoodf07}
	\end{figure*}
	
	We next address the initial density distribution of gas in UFDs. So far we have assumed that the gas follows the distribution of dark matter. However, we expect the gas to have a more concentrated density profile than the collisionless dark matter, especially during starburst events, when rapid gas transfer to the central regions fuels star formation \citep{2001AJ....122..121V,2015MNRAS.450.3886M}. On the other hand, gas can transfer energy to the dark matter via dynamical friction and thus modify the dark matter density profile, for example transforming cusps to cores \citep{2015MNRAS.446.1820N}. To test the importance of the density distribution, we explore here an extreme case where the gas density follows a NFW profile with a concentration parameter of $c=20$ as a conservative high value. This assumption implies larger gas densities, and as a result, assuming that star-formation follows the gas distribution, leads to larger retainment fractions. Since the retainment fraction in our canonical model is already quite large ($\xi\approx 0.85$), increasing the concentration does not introduce a significant change to the likelihood analysis performed in \S \ref{sec:likelihood}. Re-doing the analysis, with a highly concentrated NFW model for the density distribution and assuming SNe driven outflows, results in  $6\times 10^{-4}M_{\odot}\lesssim R_{rp/SN}\lesssim  1.5\times 10^{-2}$, $1.5\times 10^{-6}\lesssim m_{Eu}\lesssim 2.5\times 10^{-5} M_{\odot}$. These results are very similar to the results for the canonical gas density distribution model assumed in this paper, confirming that the results presented here are largely independent on the unknown density profile. The cumulative distribution of [Fe/H],[Eu/H] for a Ret II - like galaxy are shown in Fig. \ref{fig:likelihoodNFW}, in comparison to the canonical density model with SNe driven outflows.
	
	\begin{figure*}
		\centering
		\includegraphics[scale=0.39]{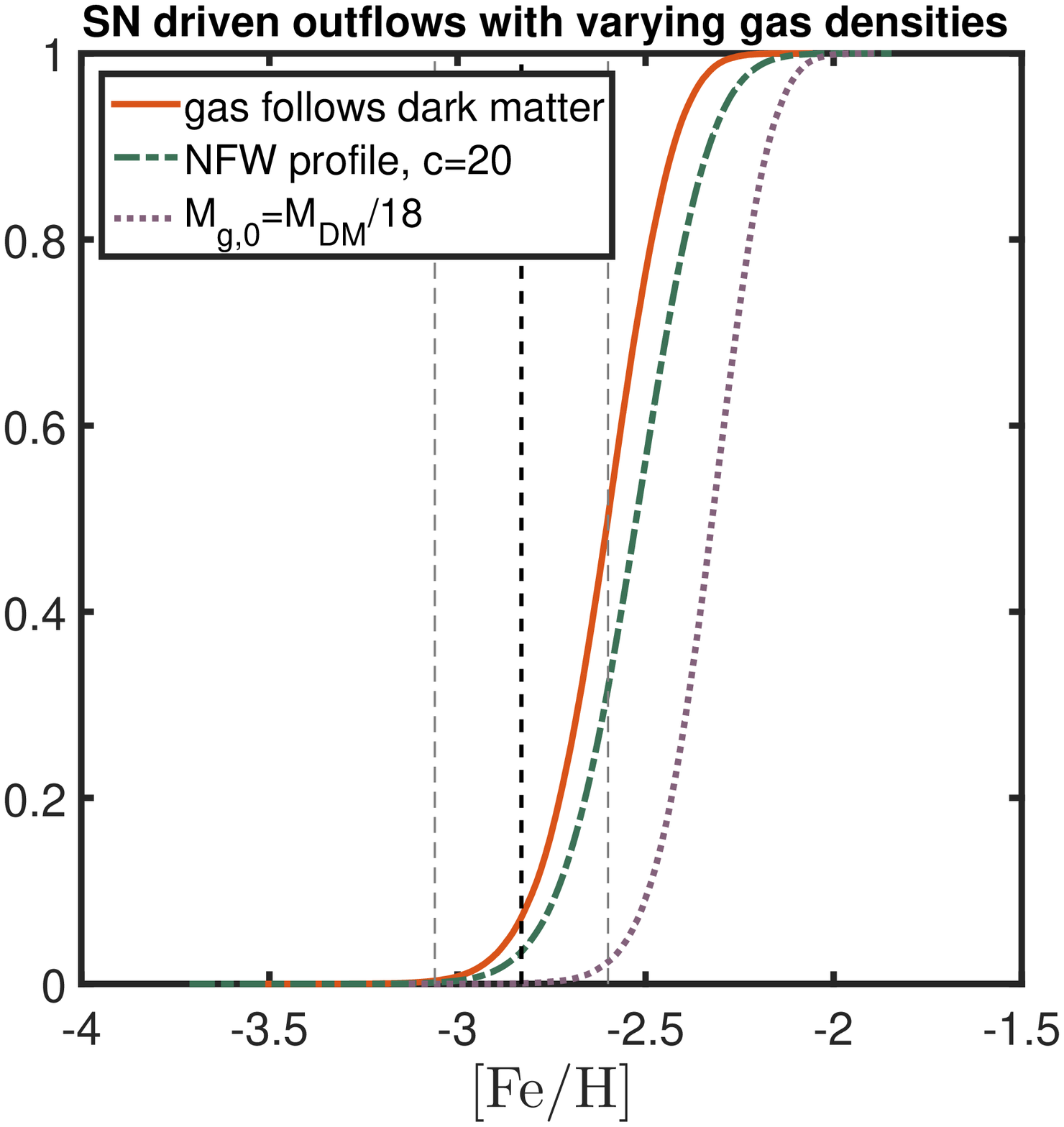}
		\includegraphics[scale=0.39]{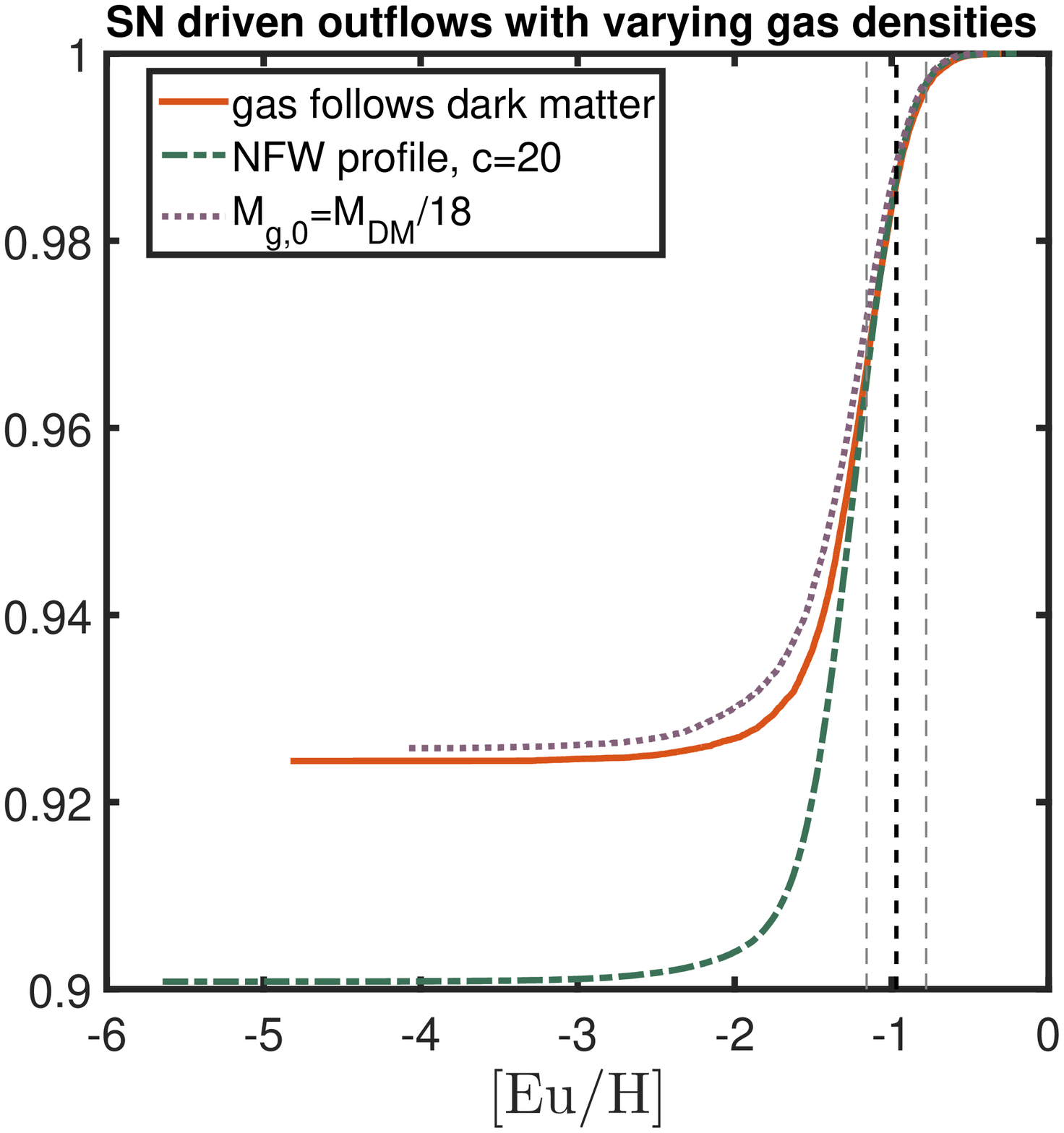}
		\caption
		{\small Cumulative distribution of [Fe/H] and [Eu/H] in an ensemble of galaxies with the same observed properties as Ret II and assuming gas loss through SNe. We compare the canonical density distribution studied in this paper in which the gas follows the dark matter density (solid lines) to an NFW profile with a concentration parameter $c=20$ (dot-dashed lines) and a profile similar to the first but with an initial gas mass lower by a facto of 3.}
		\label{fig:likelihoodNFW}
	\end{figure*}
	
	Finally, we test the dependence of the model on the initial gas mass. Throughout this work we have assumed that the initial gas mass is 1/6 of the dark matter mass, in accordance with the universal ratio between baryonic and dark matter. Since we only observe dwarf galaxies at late stages of their evolution, after significant mass loss has occurred,  the initial gas mass is not strongly constrained by observations.   Indeed, starburst dwarf galaxies, which are the most relevant for the present study have, in general, higher baryonic mass ratio \citep{2012A&A...544A.145L} than the general dwarf population \citep{2010ApJ...708L..14M}. As an illustration, we consider the effects of decreasing the initial gas mass by a factor of 3 compared to the canonical case (i.e. 1/18 of the dark matter mass). Reducing the initial gas mass, while leaving other parameters fixed to the canonical model, leads to a larger fraction of gas retained by a UFD after a given SN, as is clear from Eq. \ref{eq:whatisf}. At the same time lower gas mass, implies that iron and $r$-process elements are more easily expelled from these galaxies, and more importantly, that they mix with smaller amounts of gas mass. This affects mainly the distribution of iron abundances in this model (Fig.  \ref{fig:likelihoodNFW}) which are larger in comparison with the canonical model. Although the best fit parameters for $m_{Eu}, R_{rp/SN}$ are not strongly affected by this model, reducing the initial gas mass by a factor of three, results in a poor fit to the data. In particular, the fraction of Ret II - like galaxies that have values of [Fe/H] which are consistent with observational limits is reduced from $\sim 0.51$ in the canonical model to $\sim 0.02$. We therefore conclude that models with significantly lower initial gas mass fraction than 1/6 of the dark matter, are disfavoured by our analysis. 
	
	\section{Discussion and Summary}
	\label{sec:discuss}
	We have re-addressed the origin of $r$-process elements in ultra faint dwarf galaxies. We have focused on the implications of different gas loss models to the abundance evolution patterns in those galaxies and in particular on how they affect the fraction of $r$-process elements retained by those galaxies.
	We find that the retainment fraction in UFDs is expected to be large $\sim 0.9$, unless either a large amount of the initial galaxy's gas is removed prior to the $r$-process event and/or unless the $r$-process explosion is extremely energetic.
	In particular, this method strongly constrains $r$-process explosions involving  very large energies ($\gtrsim 10^{52}$erg) as these will require the synthesis of significantly larger amounts of $r$-process mass as compared with direct estimates from the observed abundances that do not consider the effects of retainment. For explosions involving a significant amount of gas loss from the galaxy before the $r$-process event takes place, the situation is somewhat counter intuitive. Although the retainment fraction is reduced in this case, the amount of mass into which the $r$-process material is injected is reduced even more. Overall, this means that a somewhat {\it smaller} intrinsic amount of $r$-process mass has to be created in those events, in order to explain the same observed abundances.

	For the dwarf galaxy gas loss we assume three different models. First, a simple 'top hat' model, where all of the initial gas is removed suddenly towards the end of star formation. Second, episodic mass loss events driven by SNe, and third, a constant gas outflow, that may, for instance, be driven by an AGN \citep{Silk2017}. All three models result in comparable $r$-process rates and an amount of $r$-process mass created per event that agrees to within a factor of $\sim 2$. These results are consistent with previous estimates \citep{Beniamini2016} that did not consider retainment (and that effectively assumed a 'top hat' model of mass loss) as well as with limits from the observed binary NS merger, GW170817 \citep{2017arXiv171005443D}. Given the small number of UFDs observed so far for which $r$-process abundances or limits are available, it is not yet possible to use the observed data to select between the different models. This may however be possible in the near future, as a large number of UFD galaxies are expected to be seen by upcoming missions.

	Using our best fit models for the rate and synthesized masses, we can constrain the rarity of $r$-process enriched UFDs and the typical $r$-process and iron abundances of such galaxies. All gas loss models considered here result in a fraction of $\sim 7.5\%$ of UFD galaxies that have any $r$-process enrichment. A fraction $0.1-0.27$ of these are consistent with the specific observed [Eu/H] and [Fe/H] abundances in Ret II. This implies that Ret II is a $\sim 1-2\%$ event. Interestingly, the expected iron abundances $-3.1\lesssim$ [Fe/H]$\lesssim -2.4$ are consistent with the observation of Galactic halo stars abundances $-3.5\lesssim$ [Fe/H]$\lesssim -2$ \citep{Fields2002}, as well as with the observation that these stars are found to have large star-to-star scatter in their lighter element abundances.  Specifically, \cite{Barklem2005} find that $\sim 3\%$ of metal-poor halo stars in their sample (8 out of 253) have [Eu/Fe]>1, similar to stars observed in in Reticulum II. Both the values of [Eu/Fe] and the frequency of occurrence for such stars are naturally explained if those stars originate predominantly from UFDs. Indeed, the association of metal poor halo stars with UFDs has been suggested by various previous studies \citep{Frebel2010Natur,vandevoort2015MNRAS,ishimaru2015ApJ,Griffen2016,Macias2016}; Although see \citep{RamirezRuiz2015} for a different interpretation. These results prompt us to consider the role that UFDs play as the 'building blocks' of larger galaxies, such as the Milky Way (MW) halo stars.

	Finally, we conclude with some simple estimates. Suppose the  MW PopII is made of  $10^4$   dwarf precursors, taking $10^6\rm M_\odot$ as the  fiducial stellar mass of a building block. This gives us a prediction of spatial (mixing-dependent) variations in $r$-process as well as the mean value. The former depends on a spatial diffusion model that we defer to a later discussion (see also \cite{Hotokezaka2015}). The latter can be estimated as follows: Out of $10^4$ UFD like precursors we predict $\sim 700$ with $r$-process enrichment, each with an average enhancement of the Eu to Fe density ratio by $\sim 20-40$ solar (depending on the mass loss model, see \S \ref{sec:rareRetII}), but diluted by a factor $100/7\approx 14$  for the $10^4$ dwarfs in total, all now dissolved, that form the halo stars. So on average, the Eu to Fe density ratio is twice solar, or equivalently $<\mbox{[Eu/Fe]}>\approx 0.3$ but with spatial variations of up to 100 in the density ratio (or $\mbox{[Eu/Fe]}\approx 0.3\pm 2$), depending on the degree of dilution in the halo assembly process. Note that our model allows for mass loss to the intergalactic medium (IGM), although we assume that most of this is retained in the halo as it forms. These estimates apply to the older stellar population of the Galactic halo. At later times, mixing will gradually reduce those fluctuations. Finally, when the mean iron abundance in the Milky Way, $\langle [Fe/H]\rangle$, has reached $-1$, 1a SNe start contributing significantly to iron production \citep{Nomoto2013} and these considerations no longer apply. For a discussion of the late time chemical evolution of the Galaxy, we refer the reader to \cite{Hotokezaka2018}.

	Of course this is not necessarily correct. A simple baryon budget argument tells us that half the baryons must be lost to the IGM. This is based on simulations, which show that the only model developed to day for accounting for the observed MW-type galaxy baryonic shortfall \citep{Bregman2007} involves AGN outflows in dwarf galaxy building blocks during the MW assembly phase. This AGN feedback by intermediate-mass black holes is justified by theoretical and observational evidence \citep{Peirani2012}. X-ray observations of massive spirals only detect a small fraction of the missing baryons \citep{Bogdan2017}, and the baryons most likely are located within a few virial radii \citep{Bregman2016}, possibly in intergalactic gaseous filaments \citep{graaff2017}.
	
	Another interesting aspect of retainment considerations is pollution of the IGM by $r$-process elements. Since direct $\gamma$-ray emission from radioactive decay of $r$-process material is extremely weak, it is unfortunately not possible at the current stage to observationally constrain this material. Future observations may be able to use such emission to test the topic of retainment and probe the overall number of UFDs in the local group.
	
	\section*{Acknowledgements}
	We thank Kenta Hotokezaka, Tsvi Piran and Ian Roederer for helpful suggestions and comments on the manuscript. ID is supported by the EMERGENCE 2016 project, Sorbonne Universit\'{e}s, convention no. SU-16-R-EMR-61 (MODOG).
	\bibliographystyle{mnras}
	\bibliography{retainment}
\end{document}